\documentclass[aps,pra,superscriptaddress,amsmath,amssymb,preprintnumbers,floatfix,showpacs,11pt]{revtex4}

\usepackage{amssymb}
\usepackage{epsfig}
\usepackage{subfigure}
\usepackage{graphicx}
\begin{document}
\title{ Quantum properties of the three-mode squeezed operator: triply concurrent parametric amplifiers  }
\author{  Faisal A A El-Orany$^1$,   Azeddine Messikh$^2$, Gharib S Mahmoud $^2$, Wahiddin M. R. B. }
 \affiliation{ Cyberspace Security
Laboratory, MIMOS Berhad, Technology Park Malaysia, 57000 Kuala
Lumpur, Malaysia}

 \affiliation{
International Islamic University Malaysia, P.O. Box 10, 50728
Kuala Lumpur, Malaysia }

\date{\today}

\begin{abstract}
In this paper, we study the quantum properties of the three-mode
squeezed operator. This operator is constructed from the optical
parametric oscillator based on the three concurrent $\chi^{(2)}$
nonlinearities. We give a complete treatment for  this operator
including the symmetric and asymmetric nonlinearities cases. The
action of the operator on the number and coherent states
 are studied in the framework of  squeezing, second-order correlation function,
Cauchy-Schwartz inequality and single-mode quasiprobability
function. The nonclassical effects are remarkable in all these
quantities. We show that the nonclassical effects generated by the
asymmetric case--for certain values of the system parameters--are
greater than those of the symmetric one. This reflects the
important role for the asymmetry in the system. Moreover, the
system can generate different types of the Schr\"{o}dinger-cat
states.

\end{abstract}

  \maketitle

\section{Introduction}
Squeezed light fulfils  the uncertainty relation and has less
noise than the coherent light in the one of the field quadratures.
With the development of quantum information, squeezed states have
become very important tool in providing efficient techniques for
the encoding and decoding processes in the quantum cryptography
\cite{hill}.  For instance, the two-mode squeezed states are of
great interest in the framework of continuous-variable protocol
(CVP) \cite{Ekert}. In the CVP the quantum key distribution goes
as follows. The two-mode squeezed source--such as parametric down
conversion--emits two fields: one is distributed to Alice (A) and
the other to Bob (B). Alice and Bob randomly choose to measure one
of two conjugate field quadrature amplitudes. The correlation
between the results of the same quadrature measurements on Alice's
and Bob's side increases by increasing the values of the squeezing
parameter. Through a public classical channel the users
communicate their choices for the measurements. They keep only the
results when both of them measure the same quadrature and hence
the key is generated.

The generation of the multiparities squeezed entangled states is
an essential issue  in the  multiparty communication
 including a quantum teleportation
network \cite{van1}, telecloning \cite{van2}, and controlled dense
coding \cite{van3}. The $N$-mode CV entangled states have been
generated by combining $N$ single-mode squeezed light in
appropriately coupled beam splitters \cite{van1}.  The three-mode
CV entangled states have taken much interest in the literatures,
e.g., \cite{guo,olsen,fister,van1,hua}. For instance, it has been
 theoretically shown that tripartite entanglement with different
wavelengths can be generated by cascaded nonlinear interaction in
an optical parametric oscillator cavity with parametric down
conversion and sum-frequency generation \cite{guo}. Also, the
three-mode CV  states have been generated by three concurrent
$\chi^{(2)}$ nonlinearities \cite{olsen}. This has been
experimentally verified  by the observation of the triply
coincident nonlinearities in periodically poled $KTiOPO_4$
\cite{fister}. More  details about this issue  will be given in
the next section. Furthermore, the comparison between the
tripartite entanglement in the three concurrent nonlinearities and
in the three independent squeezed states mixed on the beam
splitters \cite{van1}, in the framework of the van Loock-Furusawa
inequalities \cite{Furusawa}, has been performed in \cite{olsen}.
It is worth mentioning that  the generation of macroscopic and
spatially separated three-mode entangled light for triply coupled
$\chi^{(3)}$ Kerr coupler inside a pumped optical cavity has been
discussed in \cite{hua}. It has been shown   that the bright
three-mode squeezing and full inseparable entanglement can be
established inside and outside the cavity.

Since the early days   of the quantum optics squeezing is
connected with what is so called  squeezed operator. There have
been different forms of this operator in the literatures, e.g.
\cite{single,two,abdalla,hon,faisal,barry,marce}. For example,
degenerate and non-degenerate parametric amplifiers are sources of
the single-mode \cite{single} and the two-mode \cite{two}
squeezing, respectively.  The quantum properties of the three-mode
squeezed operator (TMS), which is constructed from  two parametric
amplifiers and one frequency converter, have been demonstrated  in
\cite{faisal}. This operator can be represented by the $SU(1,1)$
Lie algebra generators \cite{faisal,barry}. Additionally, it can
be generated--under certain condition--from  bulk nonlinear
crystal in which three dynamical modes are injected  by three
beams. Another possibility for the realization is the nonlinear
directional coupler which is composed of two optical waveguides
fabricated from some nonlinear material described by the quadratic
susceptibility $\chi^{(2)}$. Finally,  the mathematical treatments
for particular type of the $n$-mode squeezed operator against
vacuum states are given in \cite{hon}.

In this paper, we treat the three concurrent parametric amplifiers
given in \cite{olsen,fister} as three-mode squeezed operator. We
 quantitatively investigate the nonclassical effects associated
 with  this operator when  acting on the three-mode
  coherent and number states. For these states
 we investigate squeezing, second-order correlation function,
Cauchy-Schwartz inequality and single-mode quasiprobability
functions. This investigation includes the symmetric (equal
nonlinearities) and asymmetric (non-equal nonlinearities) cases.
In the previous studies   the entanglement of
 the symmetric case only has been discussed \cite{olsen,fister}.
   The investigation in the current paper  is motivated by the
importance of the three concurrent parametric amplifiers in the
quantum information research \cite{olsen,fister,van1}.
Additionally, quantifying  the nonclassical effects in the quantum
systems is of fundamental interest in its own right. We prepare
the paper in the following order. In section 2 we construct the
operator and  write down  its Bogoliubov transformations. In
sections 3 and 4 we study the quadrature squeezing, the
second-order correlation function as well as the Cauchy-Schwartz
inequality, respectively. In section 5 we investigate the
single-mode quasiprobability functions. The main results are
summarized in section 6.

\section{Operator formalism}
In this section we present the operator formalism for the optical
parametric oscillator based on the three concurrent $\chi^{(2)}$
nonlinearities. We follow the technique given in
\cite{olsen,fister} to construct the Hamiltonian of the system. In
this regard, we consider three modes injected into a  nonlinear
crystal, whose susceptibility is $\chi^{(2)}$, to form three
output beams at frequencies $\omega_0, \omega_1, \omega_2$. The
interactions are selected to couple distinct polarizations.
Assuming that $x$ is the axis of the  propagation within the
crystal. The mode $\hat{b}_1$ is pumped at frequency and
polarization $(\omega_0+\omega_1,y)$ to produce the modes
$\hat{a}_1(\omega_0,z)$ and $\hat{a}_2(\omega_1,y)$. The mode
$\hat{b}_2$ is pumped at $(\omega_1+\omega_2,y)$ to produce the
modes $\hat{a}_2$ and $\hat{a}_3(\omega_2,z)$. Eventually, the
mode $\hat{b}_3$ is pumped at $(2\omega_1,z)$ to produce the modes
$\hat{a}_1$ and $\hat{a}_2$. The scheme for this interaction can
be found in \cite{olsen,fister}.  The interaction Hamiltonian for
this concurrent triple nonlinearity takes the form
\cite{olsen,fister}:

\begin{eqnarray}
\hat{H}_{int}=i\hbar(\chi_1\hat{b}_1\hat{a}_1^\dagger\hat{a}_2^\dagger+
\chi_2\hat{b}_2 \hat{a}_1^\dagger\hat{a}_3^\dagger+
\chi_3\hat{b}_3\hat{a}_2^\dagger\hat{a}_3^\dagger)+{\rm h.c.},
\label{Ham}
\end{eqnarray}
where  $\chi_j, j=1,2,3,$ represent the effective nonlinearities
and {\rm h.c.} stands for the hermitian conjugate. The unitary
operator associated with (\ref{Ham}) is:

\begin{eqnarray}
\hat{U}(t)=\exp \left(-it\frac{\hat{H}_{int}}{\hbar}\right).
\label{Ham1}
\end{eqnarray}
In the undepleted pump approximation we  set $r_j=\chi_j\langle
\hat{b}_j(0)\rangle$ as real parameters. Now we obtain the
requested squeezed operator as:
\begin{equation}\label{1}
    \hat{S}(\underline{r})=\exp[r_1(\hat{a}_1\hat{a}_2-\hat{a}_1^{\dagger}\hat{a}_2^{\dagger})+
    r_2(\hat{a}_1\hat{a}_3-\hat{a}_1^{\dagger}\hat{a}_3^{\dagger})+
    r_3(\hat{a}_2\hat{a}_3-\hat{a}_2^{\dagger}\hat{a}_3^{\dagger})],
\end{equation}
where   $(\underline{r})=(r_1,r_2,r_3)$. Throughout this paper,
the
 symmetric case means $r_1=r_2=r_3=r$, otherwise we have an asymmetric case.  It is
evident that three disentangled state can be entangled under the
action of this operator. This operator provides the following
Bogoliubov transformations:
\begin{equation}\label{2}
  \hat{S}^{\dagger}(\underline{r})\hat{a}_j\hat{S}(\underline{r})=f_1^{(j)}\hat{a}_1+
f_2^{(j)}\hat{a}^{\dagger}_1+g_1^{(j)}\hat{a}_2+
g_2^{(j)}\hat{a}^{\dagger}_2+h_1^{(j)}\hat{a}_3+
h_2^{(j)}\hat{a}^{\dagger}_3, \quad j=1,2,3
\end{equation}
where $f^{(j)}_{j'},g^{(j)}_{j'},h^{(j)}_{j'}, j'=1,2$ are
functions in term of the parameters $r_1,r_2,r_3$.  The formulae
of these functions for the asymmetric case are rather lengthy.
Nevertheless, we write down only here the explicit forms for the
symmetric  case as \cite{olsen}:
\begin{eqnarray}
\begin{array}{lr}
  f_1^{(1)}=\frac{1}{3}[2\cosh(r)+\cosh(2r)], \quad f_2^{(1)}=\frac{1}{3}[2\sinh(r)-\sinh(2r)],   \\
  g_1^{(1)}=\frac{1}{3}[-\cosh(r)+\cosh(2r)], \quad g_2^{(1)}=-\frac{1}{3}[\sinh(r)+\sinh(2r)],   \\
g_1^{(1)}=h_1^{(1)}= f_1^{(2)}= h_1^{(2)}=f_1^{(3)}=g_1^{(3)}, \\
f_1^{(1)}=g_1^{(2)}=  h_1^{(3)},\quad f_2^{(1)}=g_2^{(2)}=  h_2^{(3)},  \\
g_2^{(1)}=h_2^{(1)}= f_2^{(2)}= h_2^{(2)}=f_2^{(3)}=g_2^{(3)}.
\label{3}
\end{array}
\end{eqnarray}
Relations (\ref{2}) and (\ref{3}) will be frequently used in the
paper. For the symmetric case, the entanglement has been already
studied in terms of the van Loock-Furusawa measure \cite{olsen}.
It has been shown that the larger the value of $r$, the greater
the quantity of the entanglement in the tripartite. Moreover, the
tripartite CV entangled state created tends towards GHZ state in
the limit of infinite squeezing, but is analogous to a W state for
finite squeezing \cite{pati}. In this paper we give an
investigation for the entanglement of the asymmetric case from
different point of view. This is based on the fact that the
entanglement between different components in the system is a
direct consequence of the occurrence of the nonclassical effects
in their compound quantities and vice versa. We show that the
asymmetric case can provide amounts of the nonclassical effects
and/or entanglement greater than those of the symmetric case. Thus
 the asymmetry in the triply concurrent parametric amplifiers is
important.

 The investigation of  the operator (\ref{1}) will be given through the three-mode squeezed
coherent and number states having the forms:
 \begin{equation}\label{12}
 |\psi_n\rangle=  \hat{S}(\underline{r})|n_1,n_2,n_3\rangle,\quad
|\psi_c\rangle=\hat{S}(\underline{r})|\alpha_1,\alpha_2,\alpha_3\rangle.
\end{equation}
Three-mode squeezed vacuum states can be obtained by simply
setting  $n_j=0$ or $\alpha_j=0$ in the above expressions. In the
following sections we study the quantum properties for the states
(\ref{12}) in greater details.

\section{Quadrature Squeezing}

Squeezing  is an important phenomenon in the quantum  theory,
which can reflect the correlation  in the compound systems very
well. Precisely, squeezing can occur in combination of the quantum
mechanical systems  even if the single systems are not themselves
squeezed. In this regard the nonclassicality of the system is a
direct consequence of the entanglement.
  Squeezed  light can be measured by the homodyne
detector, in which the signal is superimposed on a strong coherent
beam of the local oscillator. Additionally, squeezing has many
applications in various areas, e.g., in quantum optics, optics
communication, quantum information theory, etc \cite{ [21]}. Thus
investigating squeezing for the quantum mechanical systems is an
essential  subject in the quantum theory.
 In this section we demonstrate different types of squeezing for the
  three-mode
squeezed  vacuum states (\ref{12}). To do so we define two
quadratures $\hat{X}$ and $\hat{Y}$, which denote the real
(electric) and imaginary (magnetic) parts, respectively, of the
radiation field as:
\begin{equation}\label{4}
\begin{array}{lr}
\hat{X}=\frac{1}{2}[\hat{a}_1+\hat{a}^{\dagger}_1
+c_1(\hat{a}_2+\hat{a}^{\dagger}_2)+c_2(\hat{a}_3+\hat{a}^{\dagger}_3)],\\
\hat{Y}=\frac{1}{2i}[\hat{a}_1-\hat{a}^{\dagger}_1
+c_1(\hat{a}_2-\hat{a}^{\dagger}_2)+c_2(\hat{a}_3-\hat{a}^{\dagger}_3)],
\end{array}
\end{equation}
where $c_1, c_2$ are $c$-numbers take the values $0$ or $1$  to
yield  single-mode, two-mode and three-mode  squeezing. These two
operators, $\hat{X}$ and $\hat{Y}$, satisfy the following
commutation relation:
\begin{equation}
[\hat{X},\hat{Y}]=iC,
\end{equation}
 where  $C=(1+c_1^2+c_2^2)/2$.
It is said that the system is able to generate squeezing  in the
$x$- or $y$-quadrature if
\begin{eqnarray}
S_x&=&\frac{2\langle(\Delta\hat{X})^2\rangle-C}{C}<0,
\nonumber\\
&&{\rm or}\\
S_y&=&\frac{2\langle(\Delta\hat{Y})^2\rangle-C}{C}<0, \nonumber
\end{eqnarray}
where $\langle(\Delta\hat{X})^2\rangle=\langle\hat{X}^2\rangle-
\langle\hat{X}\rangle^2$ is the variance.
 Maximum squeezing occurs when $S_x=-1$ or $S_y=-1$.

 For the symmetric case, one can easily deduce the following expressions:
\begin{eqnarray}
\begin{array}{lr}
  S_x=\frac{1}{3(1+c_1^2+c_2^2)}\{(1+c_1^2+c_2^2)[2\exp(2r)+\exp(-4r)-3]
  +2(c_1+c_2+c_1c_2)[\exp(-4r)-\exp(2r)]\},
  \\
  \\
  S_y=\frac{1}{3(1+c_1^2+c_2^2)}\{(1+c_1^2+c_2^2)[2\exp(-2r)+\exp(4r)-3]
  +2(c_1+c_2+c_1c_2)[\exp(4r)-\exp(-2r)]\}.
\label{5}
\end{array}
\end{eqnarray}
For the single-mode case, $c_1=c_2=0$, the expressions (\ref{5})
reduce to:
\begin{eqnarray}\label{6}
\begin{array}{lr}
S_x=\frac{1}{3}[2\exp(2r)+\exp(-4r)-3],
\\
S_y=\frac{1}{3}[2\exp(-2r)+\exp(4r)-3].
\end{array}
\end{eqnarray}
It is evident that the system cannot generate single-mode
squeezing.   This fact is valid for the asymmetric case, too. For
the two-mode case, $c_1=1, c_2=0$, i.e. first-second mode
squeezing, we obtain
\begin{equation}\label{7}
\begin{array}{lr}
S_x=\frac{1}{3}[\exp(2r)+2\exp(-4r)-3],
\\
S_y=\frac{1}{3}[\exp(-2r)+2\exp(4r)-3].
\end{array}
\end{equation}
Squeezing can be generated in the $x$-component only with a
maximum value at $r=\ln(2)/3$. Also the maximum squeezing, i.e.
$S_x=-1$, cannot be established in this case. This is in contrast
with the
 two-mode squeezed
operator \cite{two} for which  $S_x=-1$ for large $r$. Roughly
speaking, the quantum correlation in this system  decreases the
squeezing, which can be involved in the one of  the bipartites.
Finally, for three-mode case, $c_1=c_2=1$, we have
\begin{equation}\label{8}
S_x=\exp(-4r)-1,\quad S_y=\exp(4r)-1.
\end{equation}
Squeezing can be generated in the $x$-component only  for $r>0$.
Squeezing reaches its maximum value for large values of $r$. The
origin of the occurrence squeezing in (\ref{8}) is in the strong
correlation among the components  of the system. Moreover, the
amount of the produced squeezing is two (four) times greater than
that of the two-mode \cite{two} (single-mode \cite{single})
squeezed operator  for certain  values of $r$.
\begin{figure}[h]%
  \centering
  \subfigure[]{\includegraphics[width=8cm]{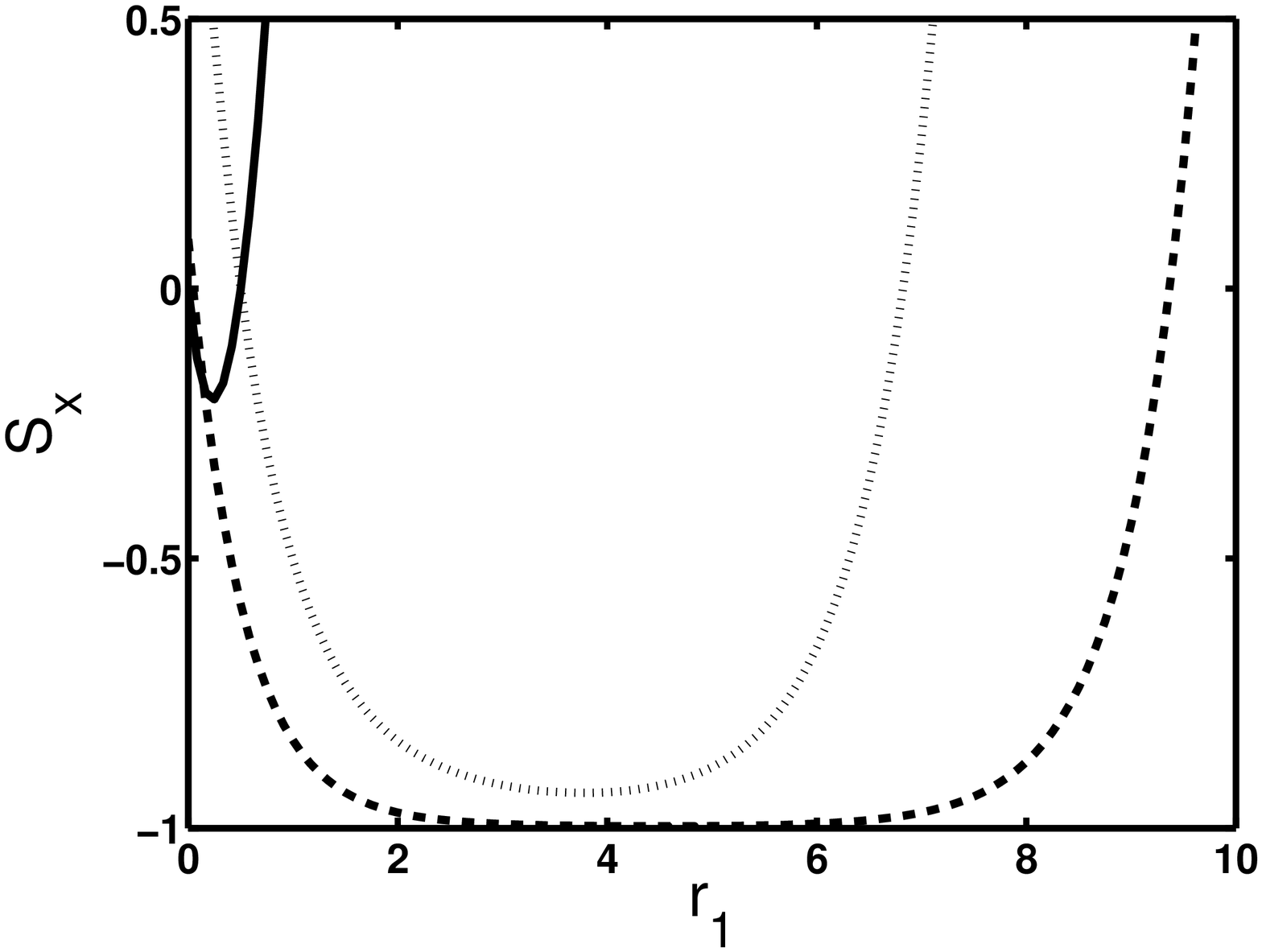}}
 \subfigure[]{\includegraphics[width=8cm]{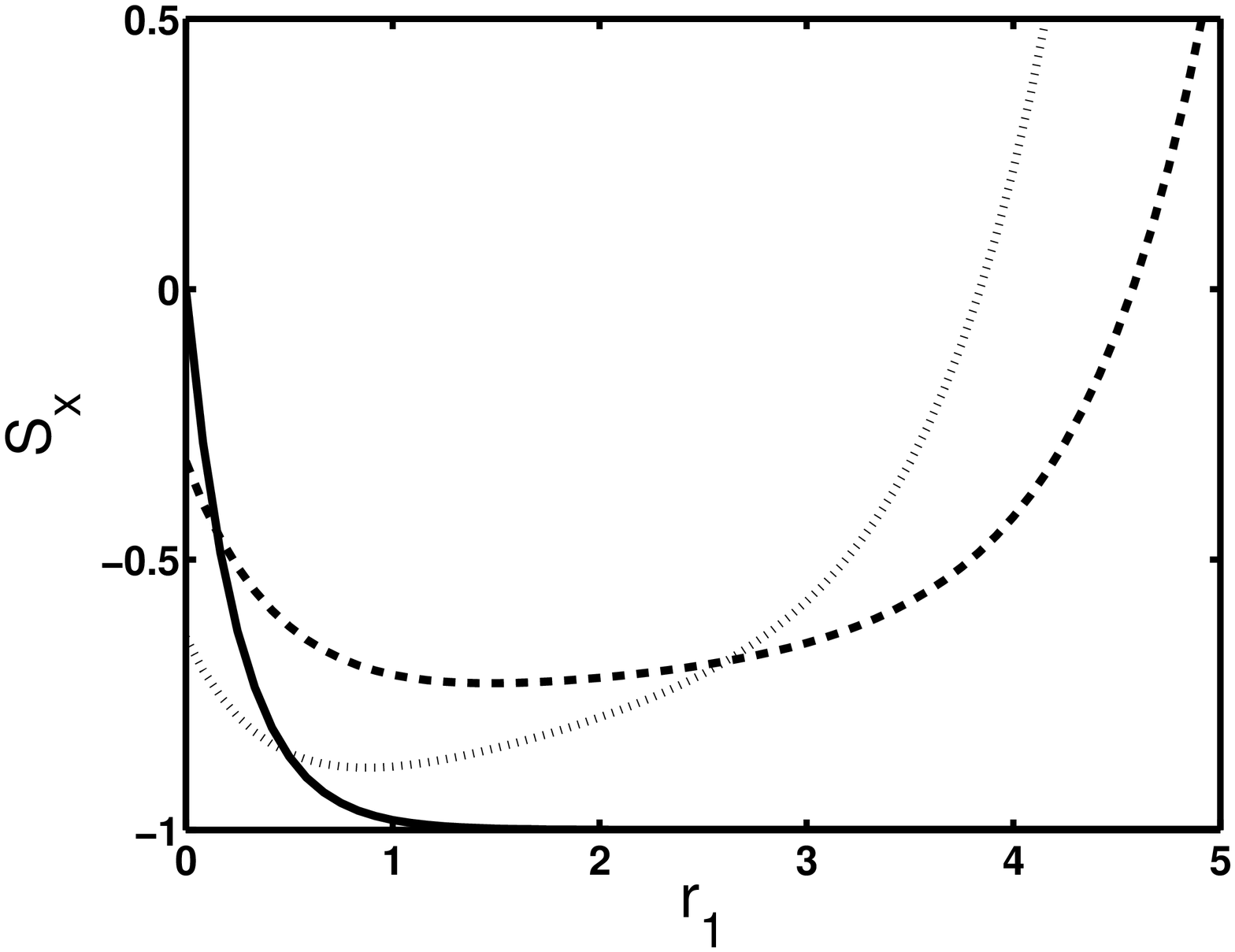}}
   \caption{
Two-mode (a) and three-mode (b) squeezing against $r_1$ for
three-mode squeezed vacuum states. Solid, dashed and dotted curves
are given for $(r_2,r_3)=(r_1,r_1), (0.1,0.2)$ and $ (0.4,0.6)$,
respectively. }
  \label{fig1}
\end{figure}

In Figs. \ref{fig1}(a) and (b) we plot the squeezing parameter
$S_x$ against $r_1$ for two- and three-mode squeezing,
respectively. We found that squeezing is not remarkable in $S_y$.
The solid curve is plotted for the symmetric case. The two-mode
squeezing is given for the first-second mode system. We start the
discussion with the two-mode case (Fig. \ref{fig1}(a)). From the
solid curve, squeezing is gradually generated as $r_1$ increases
providing  its maximum value $S_x=-0.206$ at $r=\ln(2)/3=0.231$,
then it reduces smoothly and eventually vanishes, i.e. $S_x\geq
0$,  at $r\geq \ln(1+\sqrt{3})/2=0.5025$. For the asymmetric case,
squeezing increases gradually to be maximum $S_x=-1$ over a
certain range of $r_1$, then rapidly decreases and vanishes (see
the dashed and dotted curves). This can be understood as follows.
When the values of $r_2$ and $r_3$ are relatively small, the main
contribution in the system is related to the first parametric
amplifier. Thus the system behaves  as the conventional two-mode
squeezed operator for a certain range of $r_1$. This remark is
noticeable
 when we compare the dotted curve to the dashed one. Generally,
when the values of $r_2$ and $r_3$ increase, the degradation of
the squeezing increases, too. Furthermore, in the range of $r_1$
for which $S_x=-1$, the entanglement in the bipartite $(1,2)$ is
maximum, however, this is not the case for the other bipartities.
This is connected with the  fact: the quantum entanglement cannot
be equally distributed among many different objects in the system.
Comparison among different curves in Fig. 1(a) shows that the
asymmetric case can provide amounts of squeezing much greater than
those of the symmetric case.  Now, we draw the attention to the
three-mode squeezing, which is displayed
 in Fig. \ref{fig1}(b). For the symmetric case, $S_x$ exhibits
squeezing for $r> 0$, which monotonically increases providing
maximum value for large  $r_1$, as we discussed above. This is in
a good agreement with the fact that  the symmetric case exhibits
genuine tripartite entanglement for large values of $r_1$
\cite{cont}. For the asymmetric case, the curves show initially
squeezing, which reaches its maximum by increasing $r_1$, then it
 gradually  decreases and vanishes for large values of $r_1$. The
greater the values of  $r_2, r_3$ the higher the values of the
maximum squeezing in $S_x$ and the shorter the range of $r_1$ over
which squeezing  occurs (compare dotted and dashed curves in Fig.
1(b)). This situation is the inverse of that of the two-mode case
(compare Fig. \ref{fig1}(a) to (b)). Trivial remark, for small
values of $r_1$ the amounts of squeezing produced by the symmetric
case are smaller than those by the asymmetric one. We can conclude
that for the asymmetric case the entanglement in the tripartite
may be destroyed for large $r_1$, where  $S_x>0$. Of course this
 is sensitive to the
values of $r_2, r_3$.  Conversely, the amounts of the entanglement
between different bipartites in the system for the asymmetric case
can be much greater than those of the symmetric one for particular
choice of the parameters. The final remark, the amounts of
squeezing produced by the operator (\ref{1}) are greater than
those generated by the TMS \cite{faisal}.


\section{Second-order correlation function and Cauchy-Schwartz inequality}
In this section we study  the second-order correlation function
and the Cauchy-Schwartz inequality for the states (\ref{12}).
These two quantities are  useful for getting information on the
correlations between different components in the system. In
contrast to the quadrature squeezing, these quantities are not
phase dependent and are therefore related to the particle nature
of the field.  We start with the single-mode second-order
correlation function, which for the $j$th mode  is defined as:
\begin{equation} \label{second1}
g_j^{(2)}(0)=\frac{\langle\hat{a}_j^{\dagger2}\hat{a}^2_j\rangle}
{\langle{\hat{a}^\dagger_j\hat{a}_j}\rangle^2}-1,
\end{equation}
where $g_j^{(2)}(0)=0$ for Poissonian statistics (standard case),
$g_j^{(2)}(0) < 0$ for sub-Poissonian statistics (nonclassical
effects) and $g_j^{(2)}(0) > 0$ for super-Poissonian statistics
(classical effects). The second-order correlation function can be
measured by a set of two detectors \cite{[24]}, e.g. the standard
Hanbury Brown–Twiss coincidence arrangement. Furthermore, the
sub-Poissonian light has been realized   in the resonance
fluorescence from a two-level atom driven by a resonant laser
field \cite{flour}. We have found that three-mode squeezed
coherent state cannot yield  sub-Poissonian statistics. Thus we
strict the study here to  $g_1^{(2)}(0)$ of the three-mode
squeezed number states.  Form (\ref{2}) and (\ref{12}), one can
obtain the following moments:
\begin{figure}[h]%
  \centering
\subfigure[]{\includegraphics[width=8cm]{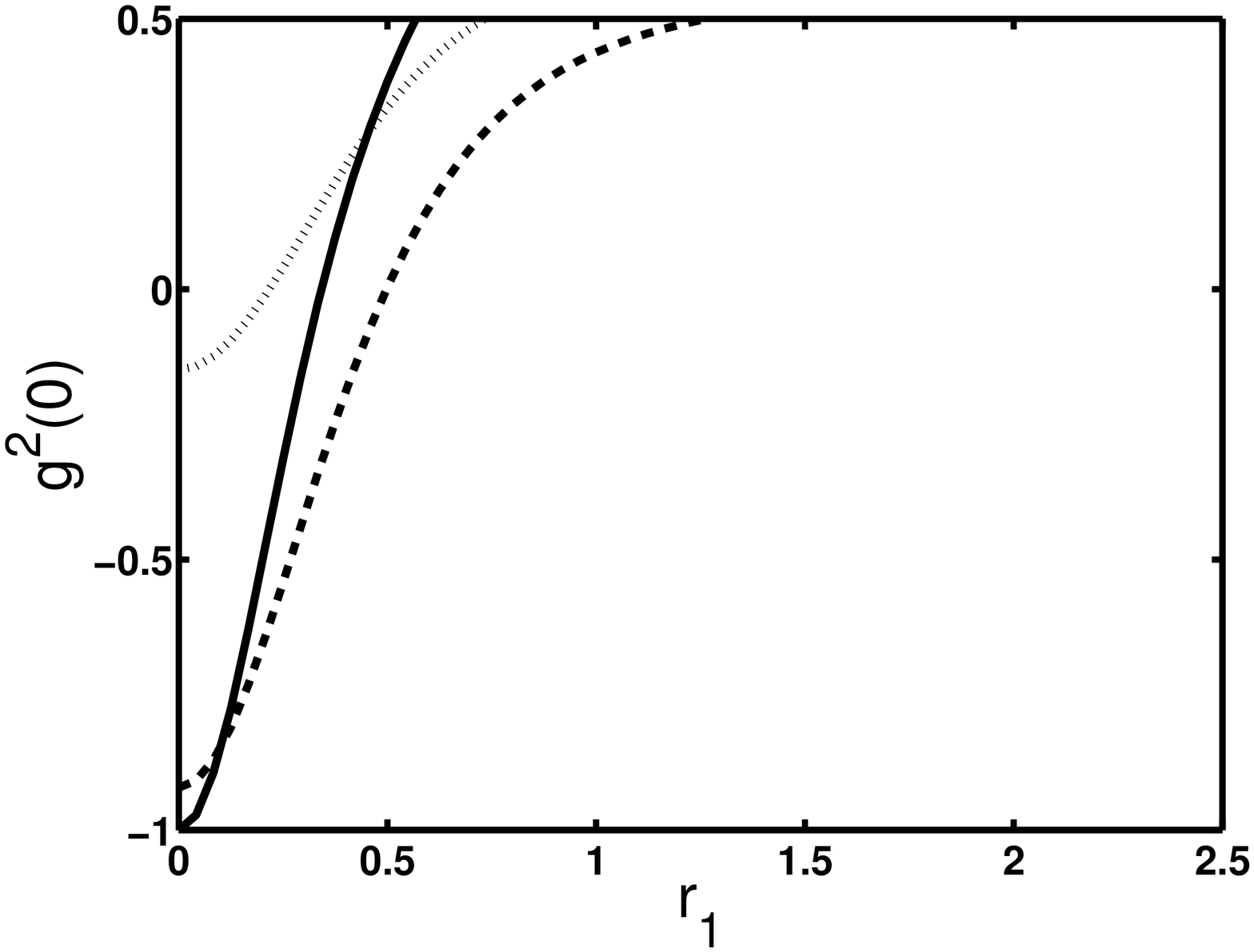}}
 \subfigure[]{\includegraphics[width=8cm]{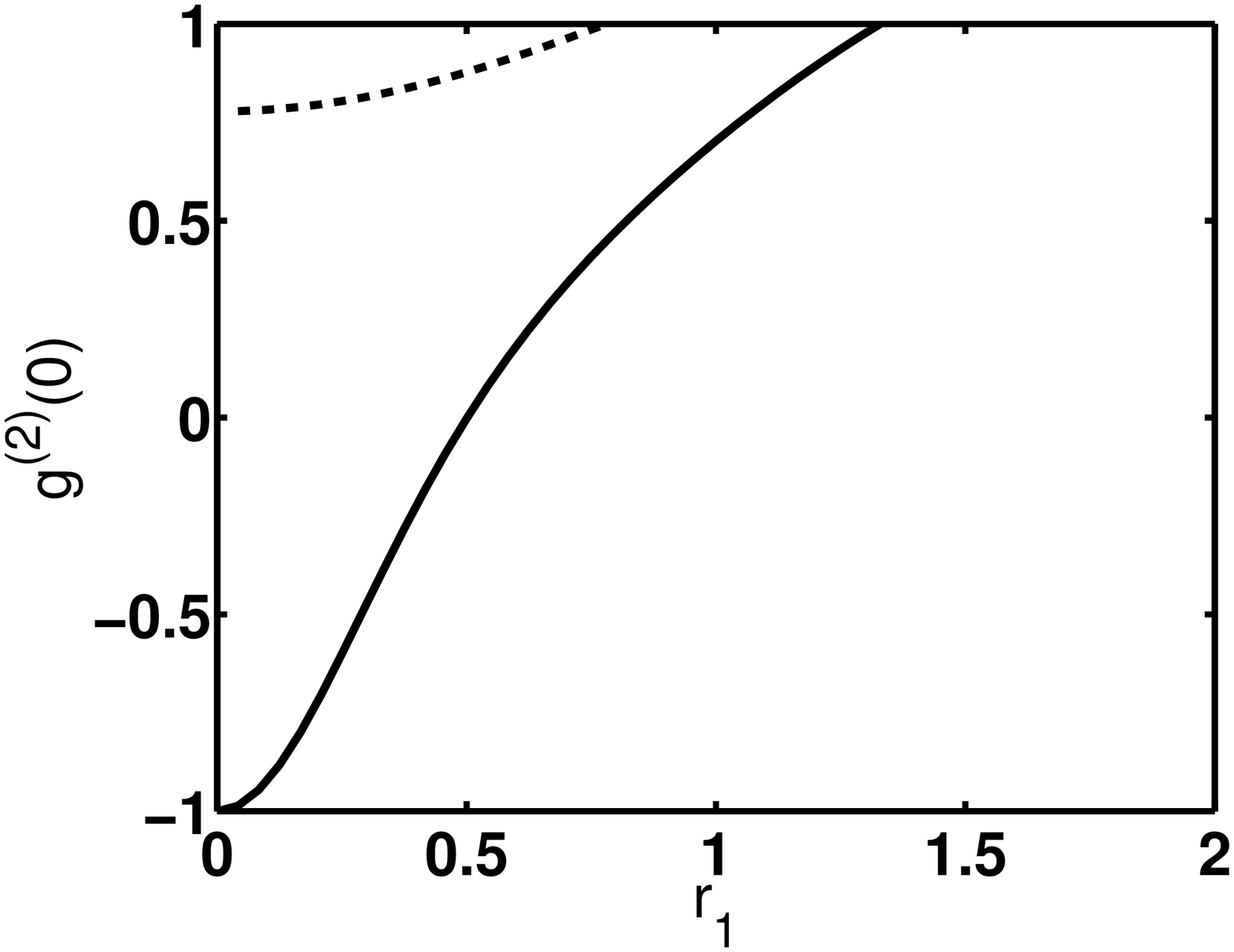}}%
  \caption{Second-order correlation of the first mode  against  $r_1$.
   (a) $(n_1,n_2,n_3)=(1,1,1)$,
 $(r_2,r_3)=(r_1,r_1)$ solid curve, $(0.1,0.2)$ dashed curve,
and $(0.4,0.6)$ dotted curve. (b) $(r_2,r_3)=(r_1,r_1)$,
$(n_1,n_2,n_3)=(1,0,0)$ solid curve and $(0,1,0)$ dashed curve.}
  \label{fig2}
\end{figure}
\begin{eqnarray}
\begin{array}{lr}
\langle{\hat{a}_1^\dagger\hat{a}_1}\rangle = n_1
f_1^{(1)2}+(n_1+1) f_2^{(1)2} +n_2 g_1^{(1)2}+(n_2+1) g_2^{(1)2}+
n_3 h_1^{(1)2}+(n_3+1) h_2^{(1)2},\\
\\
\langle{\hat{a}_1^{\dagger 2}\hat{a}_1^2}\rangle = n_1(n_1-1)
f_1^{(1)4}+(n_1+1)(n_1+2) f_2^{(1)4} +(2n_1+1)^2
f_1^{(1)2}f_2^{(1)2}\\
\\
+n_2(n_2-1) g_1^{(1)4}+(n_2+1)(n_2+2) g_2^{(1)4} +(2n_2+1)^2
g_1^{(1)2}g_2^{(1)2}\\
\\
+n_3(n_3-1) h_1^{(1)4}+(n_3+1)(n_3+2) h_2^{(1)4} +(2n_3+1)^2
h_1^{(1)2}h_2^{(1)2}\\
\\
+ (2n_1+1) f_1^{(1)}f_2^{(1)}[2(2n_2+1)
g_1^{(1)}g_2^{(1)}+(2n_3+1) h_1^{(1)}h_2^{(1)}]\\
\\
+(2n_3+1) h_1^{(1)}h_2^{(1)} [2(2n_2+1)
g_1^{(1)}g_2^{(1)}+(2n_1+1) f_1^{(1)}f_2^{(1)}]\\
\\
+4[n_1 f_1^{(1)2}+(n_1+1) f_2^{(1)2}][n_2 g_1^{(1)2}+(n_2+1)
g_2^{(1)2}+ n_3 h_1^{(1)2}+(n_3+1) h_2^{(1)2}]\\
\\
+4[n_2 g_1^{(1)2}+(n_2+1) g_2^{(1)2}][ n_3 h_1^{(1)2}+(n_3+1)
h_2^{(1)2}]. \label{second2}
\end{array}
\end{eqnarray}

By means of (\ref{second1}) and (\ref{second2}), the quantity
$g_1^{(2)}(0)$ is depicted in Figs. \ref{fig2} for given values of
the parameters. In Fig. \ref{fig2}(a) we present the role of the
squeezing parameters $r_j$ on the behavior of the $g_1^{(2)}(0)$.
From the solid curve (, i.e. symmetric case), one can observe that
the maximum sub-Piossonian statistics occur for relatively small
values of $r_j$. In this case, the system tends  to the Fock state
$|1\rangle$, which is a pure nonclassical state. As the values of
$r_1$ increase, the nonclassicality monotonically decreases and
completely vanishes around $r_1\simeq 0.3$. Comparison among the
curves in Fig. 2(a) shows when the values of $(r_2,r_3)$ increase,
the amounts of the sub-Poissonian statistics inherited in the
first mode decrease. This is connected with the nature of the
operator (\ref{1}), in which the behavior of the single-mode
 undergoes an amplification process caused by the various down-conversions
involved in the system. Furthermore,  particular values of the
asymmetry can enlarge the range of nonclassicality (compare the
solid curve to the dashed one). Now, we draw the attention to the
Fig. 2(b), which is given to the symmetric case . From this figure
one  realizes how  can obtain sub-Poissonian statistics from a
particular mode as an output from  the operator (\ref{1}).
Precisely, this mode should be initially prepared in the
nonclassical state. We can analytically prove this fact by
substituting $n_1=0, n_2=n_3=n$ into (\ref{second1}) and
(\ref{second2}). After minor algebra, we arrive at:
\begin{eqnarray}
\begin{array}{lr}
\langle\hat{a}_1^{\dagger2}\hat{a}^2_1\rangle-\langle{\hat{a}^\dagger_1\hat{a}_1}\rangle^2
  =f_2^{(1)4}+2n(n-1)g_1^{(1)4}+2(n+1)(n+2)g_2^{(1)4} \\
+2n(n+1)g_1^{(1)2}g_2^{(1)2}+[f_1^{(1)}f_2^{(1)}+2(2n+1)g_1^{(1)}g_2^{(1)}]^2 \\
+4 f_2^{(1)2}[ng_1^{(1)2}+(n+1)g_2^{(1)2}]\geq 0. \label{faisal1}
\end{array}
\end{eqnarray}
The final remark, the comparison between the solid curves in Figs.
2(a) and (b) shows that the nonclassical range of $r_1$ in (b) is
greater than that in (a). In other words, to enhance the
sub-Poissonian statistics in a certain mode, the other modes have
to be prepared in states close to the classical ones.
\begin{figure}[h]%
  \centering
\subfigure[]{\includegraphics[width=5cm]{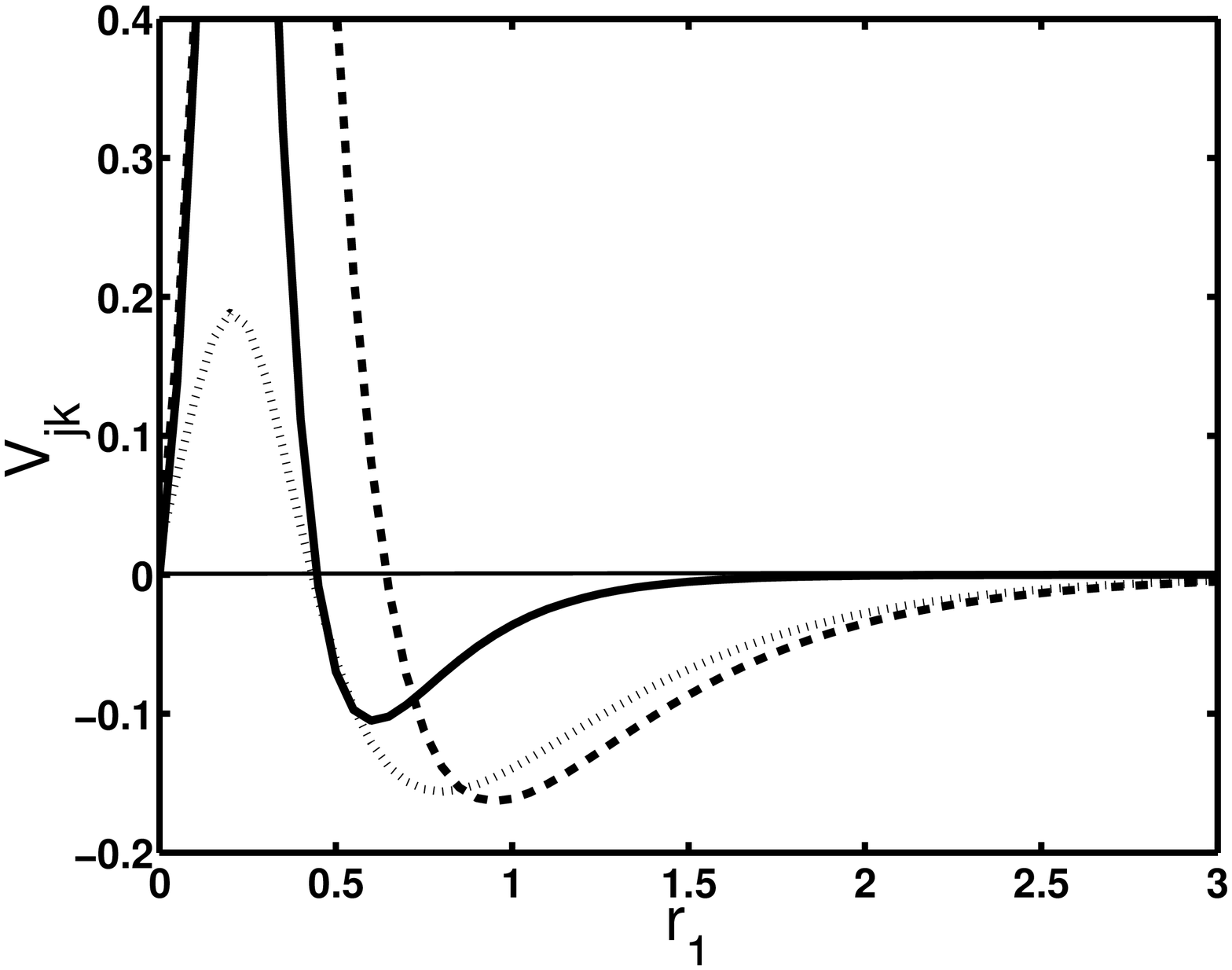}}
 \subfigure[]{\includegraphics[width=5cm]{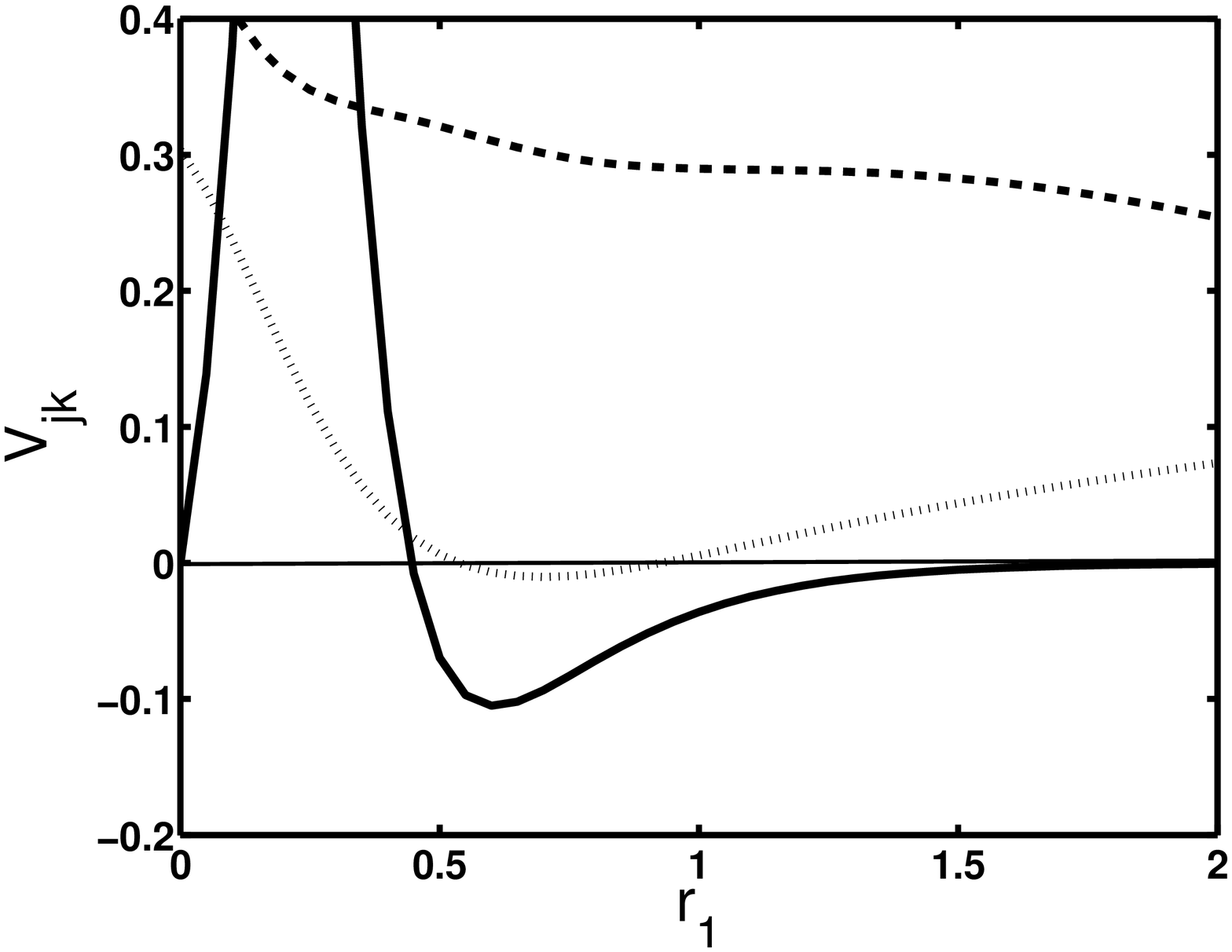}}%
\subfigure[]{\includegraphics[width=5cm]{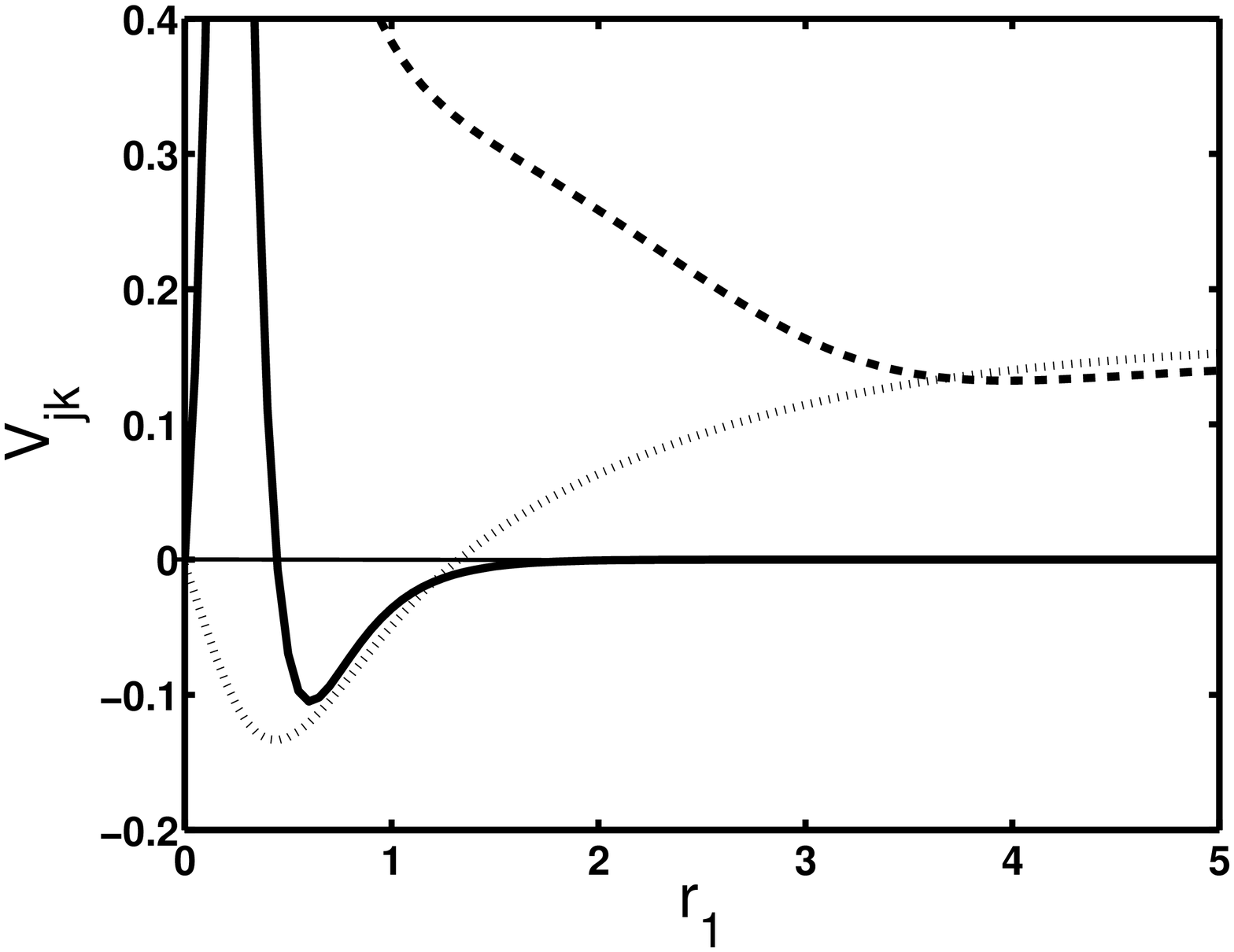}}
  \caption{The parameter $V_{jk}$ against $r_1$ for the coherent state with
$\alpha_1=\alpha_2=\alpha_3=1$, where $(j,k)=(1,2)$ (a), $(1,3)$
(b) and $(2,3)$ (c). Additionally, $(r_2,r_3)=(r_1,r_1)$ solid
curve, $(0.1,0.2)$ dashed curve  and  $(0.4,0.6)$ dotted curve. }
  \label{fig3}
\end{figure}

The violation of the classical inequalities has verified the
quantum theory.  Among these inequalities is the Cauchy-Schwarz
inequality \cite{lou}, which its violation provides information on
the intermodal correlations in the system. The first observation
of this violation  was obtained by Clauser, who used an atomic
two-photon cascade system \cite{[3]}. More recently, strong
violations using four-wave mixing have been adopted in
\cite{[4],[5]}. In addition, a frequency analysis has been used to
infer the violation of this inequality over a limited frequency
regime \cite{[11]}.
 The Cauchy-Schwarz
inequality is $V_{jk}\leq 0$, where $V_{jk}$ has the form:
\begin{eqnarray}
V_{jk}=\frac{\sqrt{\langle\hat{a}^{\dagger2}_j\hat{a}^2_j\rangle\langle\hat{a}^{\dagger2}_k\hat{a}^2_k\rangle}}
{\langle\hat{a}_j^\dagger\hat{a}_j\hat{a}_k^\dagger\hat{a}_k\rangle}-1.
\label{secont4}
\end{eqnarray}
Occurrence of the negative values in $V_{jk}$ means that the
intermodal correlation is larger than the correlation among the
photons in the same mode. This indicates a strong deviation from
the classical Cauchy-Schwarz inequality. This  is related to the
quantum mechanical features, which include pseudodistributions
instead of the true ones. In this respect, the Glauber-Sudarshan
$P$ function possesses strong quantum properties \cite{[27]}.
\begin{figure}[h]%
  \centering
  {\includegraphics[width=9cm]{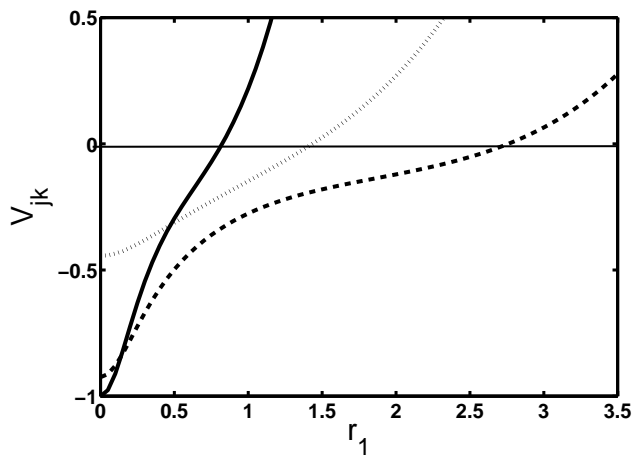}}
  \caption{
The parameter $V_{jk}$ against $r_1$ for the Fock-state case with
$n_1=n_2=n_3=1$, where $(j,k)=(1,2)$.  Additionally,
$(r_2,r_3)=(r_1,r_1)$ solid curve, $(0.1,0.2)$ dashed curve  and
$(0.4,0.6)$ dotted curve. }
  \label{fig4}
\end{figure}

We have found that the Cauchy-Schwartz inequality can be violated
for both coherent- and number-state cases. The expressions of the
different quantities in (\ref{secont4})  are too lengthy but
straightforward and hence we don't present them here. We start
with the three-mode squeezed coherent states. Information about
them is shown in Figs. 3(a), (b) and (c), for given values of the
system parameters. The negative values are remarkable in  most of
the  curves, reflecting the deviation from the classical
inequality. For the symmetric case the nonclassical correlation is
remarkable for $r_1\geq 0.4$, which increases gradually till
$r_1\simeq 0.6$ providing its maximum value, then smoothly reduces
 and eventually goes to zero, i.e. $V_{jk}\simeq 0$, for large $r_1$.
 For the
asymmetric case, the deviation from the classical inequality is
obvious, which may be smaller or greater than those in the
symmetric one based on the competition among different
nonlinearities in the system  $r_j$. In other words, e.g., the
correlation between modes $1$ and $2$ is much stronger than that
between the others only when $r_1>r_2, r_3$. This can be
understood from the structure of the operator (\ref{1}). Comparing
this behavior to that of the two-mode squeezing given in the
preceding section one can conclude that a large amount of
squeezing does not imply large violation of the inequality
\cite{carmich}. As we mentioned before:  the entanglement is a
direct consequence of the occurrence of the nonclassical effects.
As a result of this, the behavior of the two-mode squeezing and
$V_{jk}$ may provide a type of contradiction. Precisely, the
bipartite can be entangled (non-entangled) with respect to, say,
two-mode squeezing ($V_{jk}$).  This supports the fact that these
two quantities provide only a sufficient condition for
entanglement. Considering both of them we may obtain conditions
closer to the necessary and sufficient condition. A study about
this controversial issue has been already discussed for the
entanglement in a parametric converter \cite{suh}, where different
entanglement criteria leaded to different results.

In Fig. 4 we plot the parameter $V_{jk}$ for the Fock-state case.
From this figure the deviation from the classical $V_{jk}$ is
quite remarkable. For the symmetric case, maximum deviation occurs
in $V_{1,2}$ for $r=0$, monotonically decreases as $r$ evolving
and vanishes at $r \simeq 1$. Comparison among different curves in
this figure shows that the asymmetry can enlarge the range of
$r_1$ over which the deviation of the inequality occurs. Similar
behavior has been observed for $V_{2,3}$ and $V_{1,3}$ (we have
checked this fact). In conclusion, the violation of the classical
inequalities provides an explicit evidence of the quantum nature
of intermodal correlation between modes. This is not surprising,
as the entanglement is a pure quantum mechanical phenomenon that
requires a certain degree of nonclassicality either in the initial
state or in the process that governs the system.

%
\section{Quasiprobability distribution function}\label{S:sec5}
Quasiprobability distribution functions, namely, Husimi function
($Q$), Wigner function ($W$), and Glauber $P$ functions, are very
important  since they can give a global description of the
nonclassical effects in the quantum systems. These functions can
be measured by various means, e.g. photon counting experiments
\cite{[34]}, using simple experiments similar to that used in the
cavity (QED) and ion traps \cite{[35],[36]}, and homodyne
tomography \cite{[37]}. For the system under consideration, we
focus the attention here on the single-mode case, say, the first
mode for the
 three-mode squeezed number states (\ref{12}). We start with the
$s-$parameterized characteristic function $C(\zeta,s)$, which is
defined as:

\begin{equation}
C(\zeta,s)={\rm
Tr}[\hat{\rho}\exp(\zeta\hat{a}^{\dag}_1-\zeta^{*}\hat{a}_1+\frac{s}{2}|\zeta|)],\label{secf1}
\end{equation}
where $\hat{\rho}$ is the density matrix of the system under
consideration and $s$ is a parameter taking the values $0,1,-1$
corresponding to symmetrically, normally and antinormally ordered
characteristic functions, respectively.
 For three-mode squeezed number states and with the help of
the relations (\ref{2})  one can easily obtain:
\begin{eqnarray}
\begin{array}{lr}
C(\zeta,s)= \exp
[-\frac{1}{2}|\upsilon_1|^2-\frac{1}{2}|\upsilon_2|^2-\frac{1}{2}|\upsilon_3|^2+\frac{s}{2}|\zeta|^2]\\
\times {\rm L}_{n_1}( |\upsilon_1|^2) {\rm L}_{n_2}(
|\upsilon_2|^2) {\rm L}_{n_3}( |\upsilon_3|^2), \label{secf2}
\end{array}
\end{eqnarray}
where
\begin{equation}
\upsilon_1=\zeta f_1^{(1)}-\zeta^* f_2^{(1)},\quad
\upsilon_2=\zeta g_1^{(1)}-\zeta^* g_2^{(1)},\quad
\upsilon_3=\zeta h_1^{(1)}-\zeta^* h_2^{(1)} \label{secf3}
\end{equation}
and ${\rm L}_k^{\gamma}(.)$ is   the associated Laguerre
polynomial having the form:
\begin{equation}\label{reply1}
    {\rm L}_k^{\gamma}(x)=\sum\limits_{l=0}^{k}
    \frac{(\gamma+k)!(-x)^l}{(\gamma+l)!(k-l)!l!}.
\end{equation}
The $s$-parameterized quasiprobability distribution functions are
defined as
\begin{equation}
W(z,s)=\pi^{-2}\int d^{2}\zeta C(\zeta,s)\exp(z\zeta^{*}-\zeta
z^{*}), \label{secf4}
\end{equation}
where $z=x+iy$ and $s=0,1,-1$ are corresponding to $W, P, Q$
functions, respectively. On substituting  (\ref{secf2}) into
(\ref{secf4}) and applying the method of the differentiation under
the sign of integration  we can obtain the following expression:
\begin{eqnarray}
\begin{array}{lr}
W(z,s)=
\frac{1}{\pi}\sum\limits_{j',j,k=0}^{\{n_1,n_2,n_3\}}\sum\limits_{l_1,l_2=0}^{\{j',j+k\}}\left(%
\begin{array}{c}
  n_1 \\j'
\end{array}
\right)
\left(%
\begin{array}{c}
  n_2 \\k
\end{array}%
\right)
\left(%
\begin{array}{c}
  n_3 \\j
\end{array}%
\right)
\left(%
\begin{array}{c}
  j' \\l_1
\end{array}%
\right)
\left(%
\begin{array}{c}
  j+k \\l_2
\end{array}%
\right)
\frac{(-1)^{j+j'+k}}{j!j'!k!}\\
\times (f_1^{(1)2}+f_2^{(1)2})^{l_1}(g_1^{(1)2}+g_2^{(1)2})^{l_2}
(f_1^{(1)}f_2^{(1)})^{j'-l_1} (g_1^{(1)}g_2^{(1)})^{j+k-l_2}
\frac{\partial^{l_1+l_2}}{\partial b_1^{l_1+l_2}}|_{b_1=0}
\frac{\partial^{j+j'+k-l_1-l_2}}{\partial
b_2^{j+j'+k-l_1-l_2}}|_{b_2=0}\\
\times \frac{1}{\sqrt{K}}
\exp[-\frac{1}{K}(B|z|^2+(z^2+z^{*2})(\Lambda_2+b_2)],
\label{secf5}
\end{array}
\end{eqnarray}
where
\begin{eqnarray}
\begin{array}{lr}
\Lambda_1= f_1^{(1)2}+f_2^{(1)2}+g_1^{(1)2}+g_2^{(1)2}
+h_1^{(1)2}+h_2^{(1)2},\\
\\
\Lambda_2=f_1^{(1)}f_2^{(1)}+g_1^{(1)}g_2^{(1)}+h_1^{(1)}h_2^{(1)},\\
\\
B=\frac{1}{2}(\Lambda_1-s)-b_1,\quad K=B^2-(\Lambda_2+b_2)^2 .
\label{secf7}
\end{array}
\end{eqnarray}
The correlation between the modes in the system can be realized in
$W(z,s)$ as  cross terms, e.g. in $\Lambda_2$. This can give a
qualitative information about the entanglement in the system.  For
the three-mode squeezed vacuum states (, i.e., $n_1=n_2=n_3=0$)
the $W$ function  (\ref{secf5}) can be expressed as:
\begin{equation}
W(x,y,s)=\frac{1}{\pi \sqrt{\vartheta_{+}\vartheta_{-}}}
\exp[-\frac{x^2}{\vartheta_{+}}-\frac{y^2}{\vartheta_{-}}],
\label{tsecf8}
\end{equation}
where
\begin{eqnarray}
\begin{array}{lr}
\vartheta_{+}=2\langle(\Delta\hat{X}_1)^2\rangle-\frac{s}{2},\\
\\
=\frac{1}{2}[(f_1^{(1)}+f_2^{(1)})^2+(g_1^{(1)}+g_2^{(1)})^2+
(h_1^{(1)}+h_2^{(1)})^2]-\frac{s}{2},\\
\\
\vartheta_{-}=2\langle(\Delta\hat{Y}_1)^2\rangle-\frac{s}{2},\\
\\
=\frac{1}{2}[(f_1^{(1)}-f_2^{(1)})^2+(g_1^{(1)}-g_2^{(1)})^2+
(h_1^{(1)}-h_2^{(1)})^2]-\frac{s}{2}. \label{tsecf10}
\end{array}
\end{eqnarray}
From (\ref{tsecf8}) and (\ref{tsecf10}) it is evident that the
quasidistributions are Gaussians, narrowed in the $y$ direction
and expanded in the $x$ direction. Nevertheless, this does not
mean squeezing is available in this mode. Actually, this behavior
represents the thermal squeezed light, which, in this case, is a
super-classical light. Precisely, with $s=0$ the $W$ function
exhibits  stretched contour, whose area  is broader than that of
the coherent light. In this regard the phase distribution and the
photon-number distribution associated with the single-mode case
exhibits a single-peak structure for all values of $r_j$. This
peak is broader than that of the coherent state, which has the
same mean-photon number. Actually, this is a quite common property
for the multimode squeezed operators
\cite{single,two,abdalla,hon,faisal,barry,marce}. Thus the
single-mode vacuum or coherent states, as outputs from the three
concurrent amplifiers described by (\ref{1}) are not nonclassical
states. This agrees with the information given in the sections 3
and 4.

Now we consider two cases:  $(n_1,n_2,n_3)=(0,0,n_3)$ and
$(n_1,n_2,n_3)=(n_1,0,0)$.  For the first case, we investigate the
influence of the nonclassicality in the third mode on the behavior
of the first mode.  To do so we substitute $n_1=n_2=0$  in
 (\ref{secf5}) and after minor algebra we arrive at:
\begin{eqnarray}
\begin{array}{lr}
W(x,y,s)=
\frac{(-1)^{n_3}}{\pi\sqrt{(\frac{\Lambda_1-s}{2})^2-\Lambda_2^2}}(\frac{\eta_-}{\vartheta_+})^{n_3}
\exp[-\frac{x^2}{\vartheta_-}-\frac{y^2}{\vartheta_+}]\\
\\
\times
\sum\limits_{m=0}^{n_3}\left(\frac{\vartheta_+\eta_+}{\vartheta_-\eta_-}\right)^m
{\rm
L}_m^{-\frac{1}{2}}[(\frac{\eta_++\vartheta_-}{\eta_+\vartheta_-})x^2]
{\rm
L}_{n_3-m}^{-\frac{1}{2}}[(\frac{\eta_-+\vartheta_+}{\eta_-\vartheta_+})y^2]
, \label{secf10}
\end{array}
\end{eqnarray}
where
\begin{equation}
\eta_{\pm}=(h_1^{(1)}\pm h_2^{(1)})^2-\vartheta_{\mp}.
\label{secf11}
\end{equation}
One can easily check when $r_1=r_2=r_3=0$ the $W$ function
(\ref{secf10}) reduces to that of the vacuum state. The form
(\ref{secf10})  includes Laguerre polynomial, which is well known
in the literature by providing nonclassical effects in the phase
space. This indicates that the nonclasssical effects can be
transferred from one mode to another under the action of the
operator (\ref{1}). Of course the amount  of the transferred data
depends on the values of the squeezing parameters $r_j$.

\begin{figure}[h]%
  \centering
  \subfigure[]{\includegraphics[width=8cm]{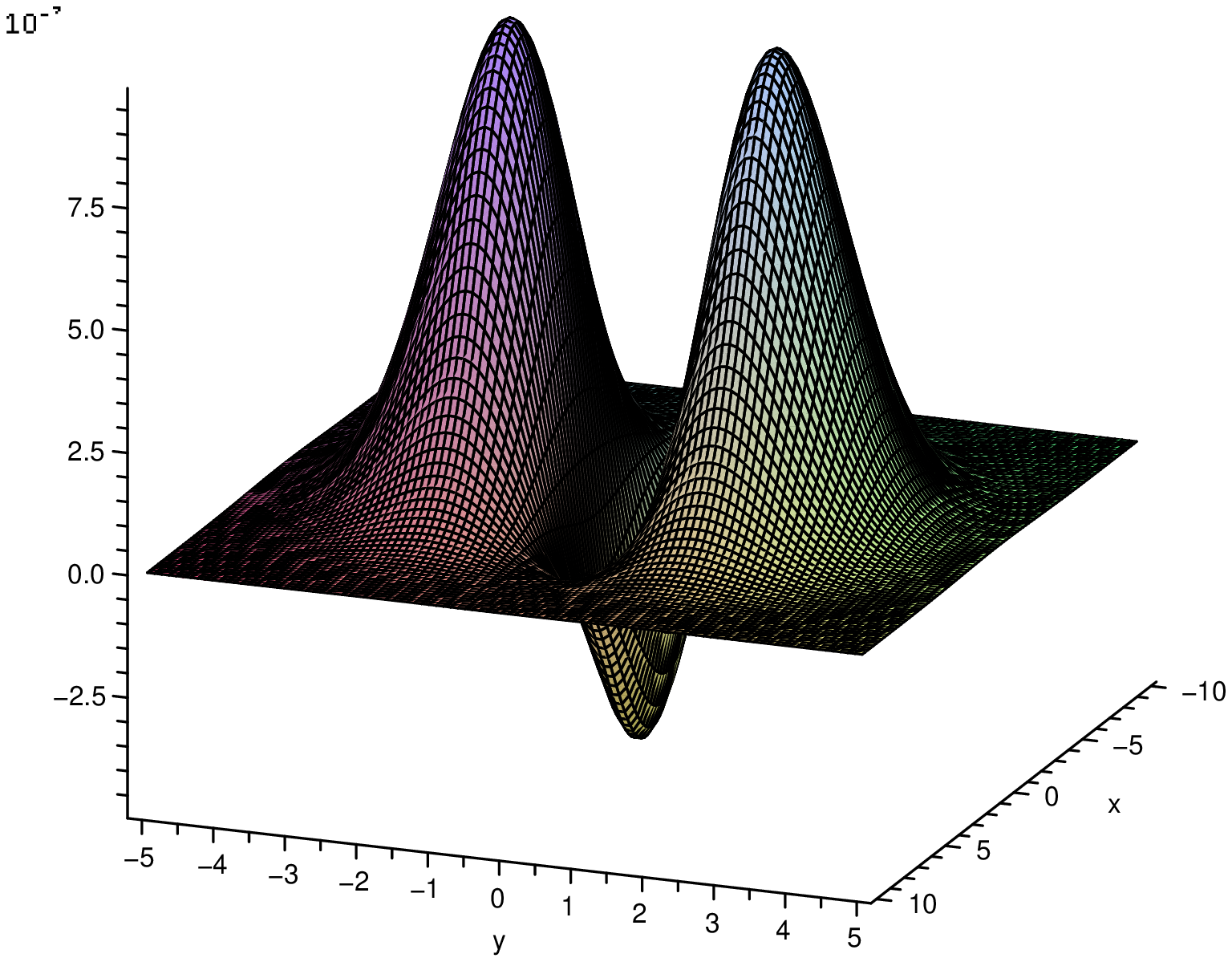}}
 \subfigure[]{\includegraphics[width=8cm]{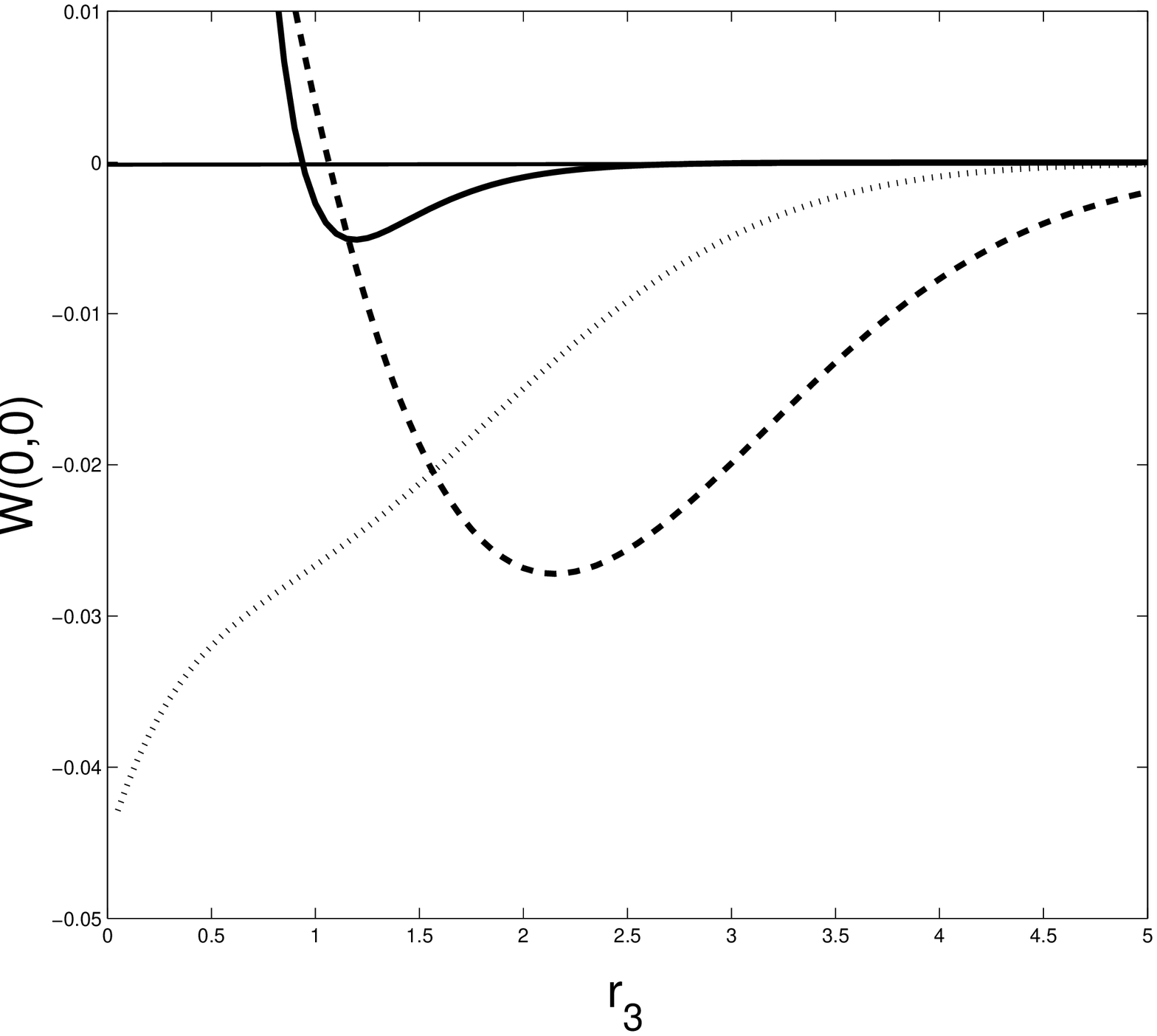}}%
  \caption{The Wigner function of the first mode (a) and the evolution of the phase space origin of the Wigner function (b).
  In (a) we use $(r_1,r_2,r_3,n_1,n_2,n_3)=(1.1,1.1,1.1,0,0,1)$. In (b) we use $(r_1,r_2,n_1,n_2,n_3)=(r_3,r_3,0,0,1)$ solid
  curve,
$(0.6,0.8,0,0,1)$ dashed curve, and $(0.6,0.8,1,0,0)$ dotted
curve. }
  \label{fig5}
\end{figure}

The second case $(n_1,0,0)$ has the same expression (\ref{secf10})
with the following transformations:
\begin{equation}
n_3\rightarrow n_1,\quad \eta_{\pm}=(f_1^{(1)}\pm f_2^{(1)})^2-
\vartheta_{\mp}. \label{secf13}
\end{equation}
Now we prove that this $W$ function tends to that of the number
state when  $r_1=r_2=r_3=0$. In this case, the transformations
(\ref{secf13}) tend to:
\begin{equation}
\eta_{\pm}=\frac{1+s}{2}, \quad \vartheta_{\pm}=\frac{1-s}{2}.
\label{secf14}
\end{equation}
Substituting these variables  in  the expression (\ref{secf10})
(with $n_3\rightarrow n_1$) we obtain:
\begin{eqnarray}
\begin{array}{lr}
W(x,y,s)= \frac{2(-1)^{n_1}}{\pi}
\frac{(1+s)^{n_1}}{(1-s)^{n_1+1}}\exp[-\frac{2(x^2+y^2)}{1-s}]
\sum\limits_{m=0}^{n_1} {\rm
L}_m^{-\frac{1}{2}}[\frac{4x^2}{1-s^2}] {\rm
L}_{n_1-m}^{-\frac{1}{2}}[\frac{4y^2}{1-s^2}]\\
\\
=\frac{2(-1)^{n_1}}{\pi}
\frac{(1+s)^{n_1}}{(1-s)^{n_1+1}}\exp[-\frac{2(x^2+y^2)}{1-s}]
 {\rm
L}_{n_1}[\frac{4(x^2+y^2)}{1-s^2}] , \label{secf15}
\end{array}
\end{eqnarray}
In (\ref{secf15}) the transition from the first line to the second
one has been done using  the identity:
\begin{equation}
\sum\limits_{n=0}^{m} {\rm L}_n^{\tau_1}(x) {\rm
L}_{m-n}^{\tau_2}(y)={\rm L}_m^{\tau_1+\tau_2+1}(x+y).
\label{secf16}
\end{equation}
The expression (\ref{secf15}) is the $s$-quasprobability
distribution for the number state, e.g. \cite{wolfgang}.
\begin{figure}
\centering \subfigure[]{\includegraphics[width=7cm]{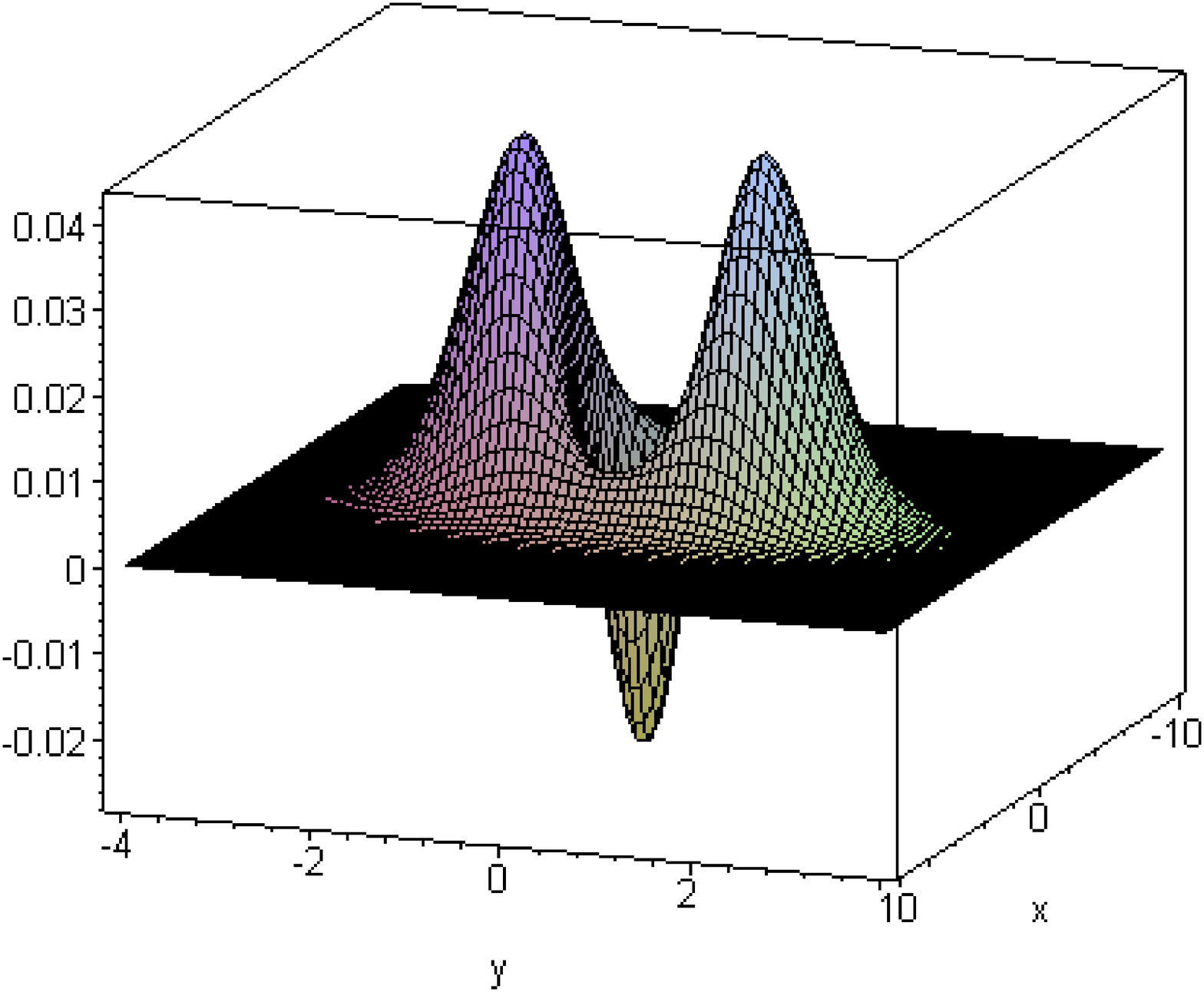}}
\subfigure[] {\includegraphics[width=7cm]{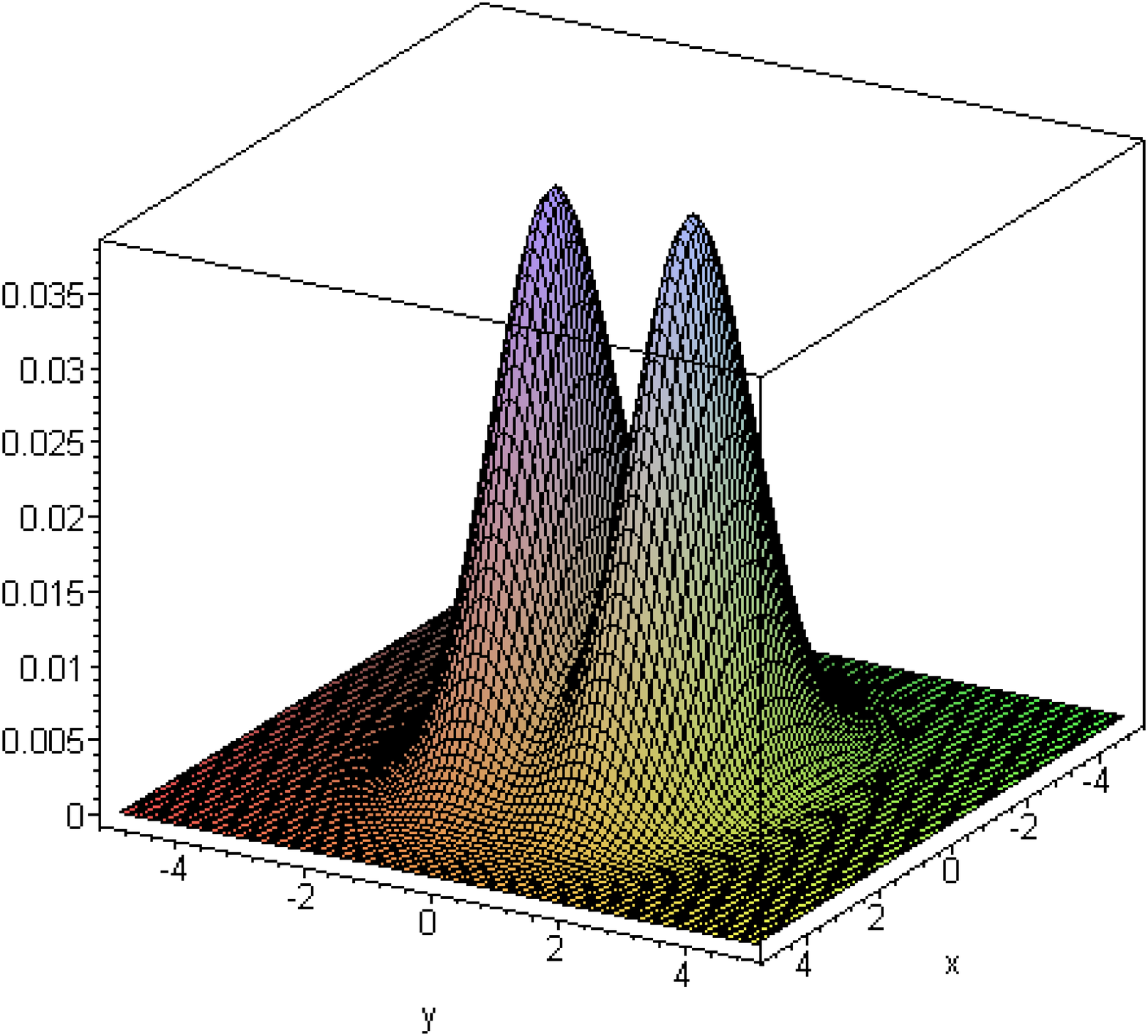}}
\subfigure[] {\includegraphics[width=7cm]{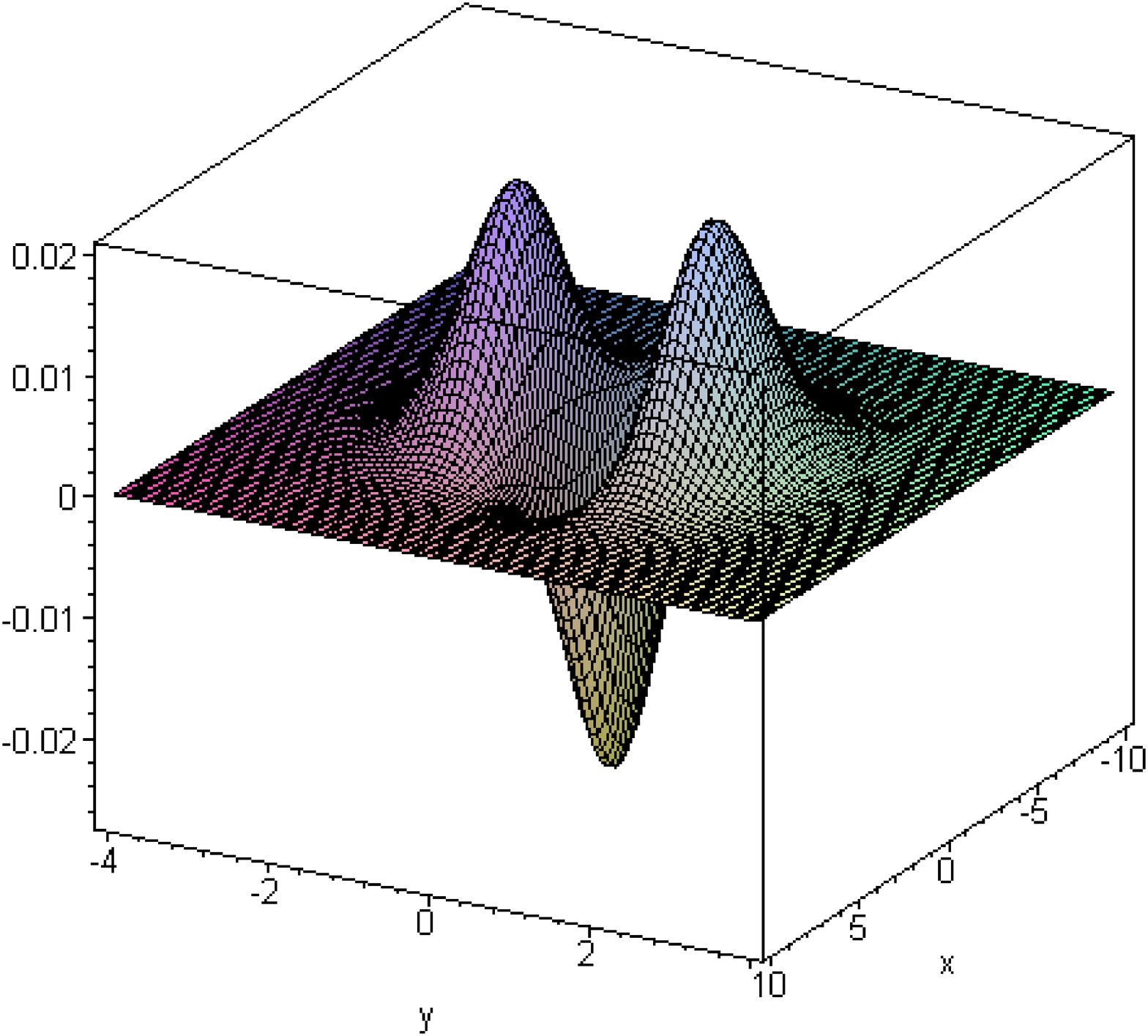}}
\subfigure[]{\includegraphics[width=7cm]{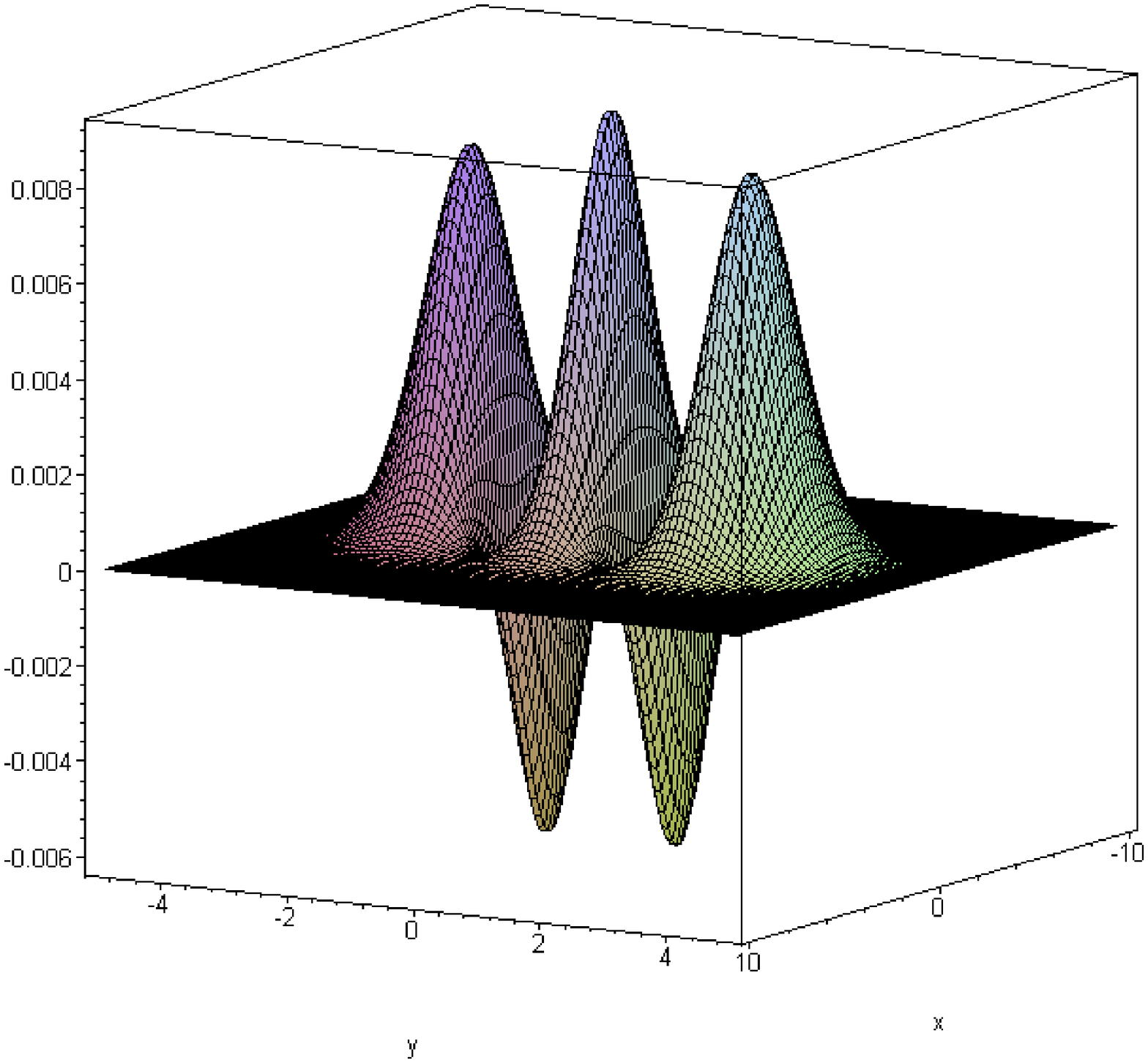}}
\caption{The Wigner function of the first mode when
$(r_1,r_2,r_3,n_1,n_2,n_3)=(0.6,0.8,0.9,1,0,0)$ (a),
$(0.6,0.8,0.9,0,0,1)$ (b),  $(0.6,0.8,2,0,0,1)$ (c) and
$(0.4,0.8,2,0,0,2)$ (d).
 } \label{fig6}
\end{figure}
We conclude this part by writing down the form of the $W$ function
at the phase space origin, which is a sensitive point for the
occurrence of the nonclassical effects. Moreover, it can simply
give visualization about the behavior of the system. Additionally,
this point  can be measured by the photon counting method
\cite{[34]}. From (\ref{secf10}) we have:
\begin{equation}
W(0,0,s)= \frac{(-1)^{n_3}}{\pi\sqrt{\vartheta_+\vartheta_-}}
(\frac{\eta_-}{\vartheta_+}+\frac{\eta_+}{\vartheta_-})^{n_3}.
\label{[28]}
\end{equation}
It is obvious that the Winger function exhibits negative values at
the phase space origin only when $n_3$ is an odd number.

In Figs. 5 and 6 we plot  the $W$ functions for the given values
of the system parameters. We start the discussion with the
symmetric case. In Fig. 5(a) we use $(n_1,n_2,n_3)=(0,0,1)$
meaning that the mode under consideration is in the vacuum state.
Thus for $r_j=0$ the $W$ function exhibits the
single-peak-Gaussian structure with a center at the phase space
origin. When $r_j\neq 0$  this behavior is completely changed,
where one can observe  a lot of the nonclassical features, e.g.
 negative values, multipeak structure  and stretching contour (see Fig. 5(a)).
  This indicates that the nonclassical effects can be transferred from one mode to the other under
 the action of the operator (\ref{1}).  In
 Fig. 5(b) we plot the "evolution" of the $W$ function given by (\ref{[28]}) against
the parameter $r\quad (r_3)$ for the
 symmetric (asymmetric) case.  The aim of this figure is to estimate
 the exact value of  the  nonlinearity  $r\quad (r_3)$ for
which  the nonclassical effects  maximally occur and/or transfer
from certain mode to the other.   For the symmetric case,  this
occurs at $r=1.2$, while the nonclassicality is completely washed
out at $r=3$. Now, we draw the attention to the asymmetric case
which is plotted in Figs. 6. Fig. 6(a) gives information on the
case $(n_1,n_2,n_3)=(1,0,0)$. The $W$ function of the Fock state
$|1\rangle$ is well known in the literatures by having inverted
peak in phase space with maximum negative values. This is related
to that this state provides maximum sub-Poissonian statistics.
Under the action of the operator (\ref{1}) these negative values
are reduced and the two-peak structure is started to be
constructed. This indicates that the system is able to generate
particular types of the Schr\"{o}dinger-cat states by controlling
the system parameters. This is really obvious  in Figs. 6(b)--(d),
which are given to the cases $n_1=n_2=0, n_3=1,2$. For instance,
from Fig. 6(b) the $W$ function provides a two-peak structure.
Nevertheless, by increasing the values of the parameter $r_3$, the
$W$ function exhibits the two Gaussian peaks and inverted negative
peak in-between indicating the occurrence of the interference in
phase space (see Fig. 6(c)). This shape is similar to that of the
odd-coherent state. Additionally, the Fig. 6(d), in which $n_3=2$,
provides the well-known shape of the $W$ function of the even
coherent state. Generally, the even and the odd
Schr\"{o}dinger-cat  states have nearly identical classical
components (, i.e. the positive peaks) and only differ in the sign
of their quantum interferences.  These are interesting results,
which show that by controlling the nonlinearity of the system and
preparing  a certain mode in the Fock state $|1\rangle$ or
$|2\rangle$ one can generate  cat states. It is worth mentioning
that the Fock state $|n\rangle$ can be prepared with very high
efficiency according to the recent experiments \cite{[48]}.
 Similar results
have been obtained from the codirectional three-mode Kerr
nonlinear coupler \cite{faisalcoupler}. Furthermore, quite
recently the construction of the cat state trapped in the cavity
in which several photons survive long enough to be repeatedly
measured is given in \cite{haroche}. In this technique, the atoms
crossing the cavity one by one are used to obtain information
about the field. We proceed, we have noted that the $W$ function
of the cases $ n_2=n_3=0, n_1=1,2$ can provide quite similar
behaviors as those of Figs. 6(c) and (d) for the same values of
$r_j$. Now, we draw the attention to the dashed and dotted curves
in the Fig. 5(b). These curves provide information on the
evolution of the $W(0,0)$ against $r_3$ for the case of Fig. 6(a)
and (c), respectively. From the dotted curve, i.e. the mode under
consideration is in the Fock state, the $W(0,0)$ exhibits the
maximum negativity at $r_3=0$, which monotonically decreases and
completely vanishes at $r_3=4$. This shows for how long  the
nonclassicality inherited in the first mode  survives based on the
intensity of the third amplifier. On the other hand, form the
dashed curve, i.e.  the mode under consideration is in the vacuum
state, the $W(0,0)$ exhibits negative values for $r_3\geq 1$,
increases rapidly to show maximum around $r_3=2$, reduces
gradually and vanishes for $r_3\geq 5$.
 This range of negativity is greater than
that of the dotted curve. This is in a good agrement  with the
behavior of the second-order correlation function.
 Finally,
comparison among the curves  in Fig. 5(b) confirms the fact:  for
certain values of the system parameters, the asymmetric case can
provide nonclassical effects greater than those of the symmetric
one.

\section{Conclusion}
In this paper we have studied the three-mode squeezed operator,
which can be implemented from the triply coincident nonlinearities
in periodically poled $KTiOPO_4$. The action of this operator on
the three-mode coherent and number states is demonstrated.  We
have studied quadrature squeezing, second-order correlation
function, Cauchy-Schwartz inequality and quasiprobability
distribution function. The obtained results can be summarized as
follows. Generally, the single-mode vacuum or coherent states, as
outputs from the three concurrent amplifiers are not nonclassical
states. The system can exhibit two-mode and three-mode squeezing.
The amount of the two-mode squeezing generated by the asymmetric
case is much greater than that of the symmetric case. Three-mode
squeezed coherent (number) states cannot (can) exhibit
sub-Poissonian statistics.
 To obtain maximum
sub-Poissonian statistics
 from a particular mode, under the action of the operator (\ref{1}),
  it must be  prepared in the nonclassical
 state and the other modes in states close to the classical ones.
We have found that the Cauchy-Schwartz inequality can be violated
for both coherent states and number states. The origin in the
violation is in the strong  quantum correlation among different
modes. For the Fock-state case, the asymmetry in the system
enhances the range of nonlinearities for which $V_{jk}$ is
nonclassical compared to that of the symmetric one. In the
framework of the quasiprobability distribution we have shown that
 the nonclassical effects can be transferred from one mode to
another under the action of the operator (\ref{1}). The amount of
transferred  nonclassicality is sensitive to the values of the
squeezing parameters.
 Interestingly, the system can generate particular
types of the Schr\"{o}dinger-cat states for certain values of the
system parameters. Generally, we have found that the nonclassical
effects generated by the operator (\ref{1}) are greater than those
obtained from the operator TMS \cite{faisal}.
 Finally,  the asymmetry in
the three concurrent nonlinearities process is important for
obtaining  significant nonclassical effects.

\section*{References}


\begin{thebibliography}{200}

\bibitem{hill}
Hillery M 2000 {\it Phys. Rev. A} {\bf 61} 022309.

\bibitem{Ekert}  Ekert A K 1991 {\it Phys. Rev. Lett.} {\bf 67} 661;
  Bennett C H,  Brassard G and   Mermin N D 1992 {\it Phys. Rev. Lett.} {\bf
68}  557.



\bibitem{van1}  Loock P V and  Braunstein S L 2000 {\it Phys. Rev. Lett.} {\bf 84}
3482.
\bibitem{van2}  Loock P V and  Braunstein S L 2001 {\it Phys. Rev. Lett.} {\bf
87} 247901.

\bibitem{van3}  Jing J,  Zhang J,  Yang Y, Zhao F,
Xie C and Peng K 2003 {\it Phys. Rev. Lett.} {\bf 90} 167903.




\bibitem{guo}
 Guo J,  Zou H,  Zhai Z,  Zhang J and Jiangrui
2005 {\it Phys. Rev. A} {\bf 71} 034305.

\bibitem{olsen} Olsen M K, Bradley A S and Reid M D 2006 {\it J Phy
B: At. Mol. Opt. Phys.} {\bf 39 } 2515;  Bradley A S, Olsen M K,
Pfister O and  Pooser R C 2005 {\it Phys. Rev. A} {\bf 72} 053805.




\bibitem{fister}  Pfister O,  Feng S,  Jennings G,  Pooser R C and  Xie D 2004 {\it Phys.
Rev. A} {\bf 70} 020302(R); Pooser R C  and  Pfister O 2005 {\it
Opt. Lett.} {\bf 30} 2635.


\bibitem{hua}
 Tan H-T; Li G-X and  Zhu S-Y 2007 {\it Phys. Rev. A} {\bf 75}
063815.

\bibitem{Furusawa}Loock P V and Furusawa A 2000 {\it Phys. Rev. A.} {\bf 67}
052315.

\bibitem{single}Yuen H P 1976 {\it Phys. Rev. A} {\bf 13} 2226;
 Caves C M 1981 {\it Phys. Rev. D.} {\bf 23} 1693.

\bibitem{two} Barnett S M and Knight P L 1985 {\it J Opt. Soc. Am.
B} {\bf 2} 467; Barnett S M and Knight P L 1987 {\it J. Mod. Opt.}
{\bf 34} 841; Gilles L and Knight P L 1992 {\it J. Mod. Opt.} {\bf
39} 1411.

\bibitem{abdalla}  Abdalla M S 1992 {\it J. Mod. Opt.} {\bf 39} 771; {\it ibid.} {\bf 39} 1067; 1993 {\it
ibid.} {\bf 40} 441; {\it ibid.} {\bf 40} 1369;  Abdalla M S and
Obada A-S F 2000 {\it Int. J. Mod. Phys.} {\bf 14} 1105.
\bibitem{faisal}
 Abdalla M S,  El-Orany F A A and  Pe\v{r}ina J 2001 {\it Eur.
Phys. J. D.} {\bf 13}, 423.

\bibitem{barry}  Shaterzadeh-Yazdi Z,  Turner P S and
Sanders B C 2008 {\it J. Phys. A: Math. Theor.} {\bf 41} 055309.


\bibitem{hon}  Hu L-Y and  Fan H-Y 2009 {\it EPL} {\bf 85}  60001.



\bibitem{marce}
 Marchiolli M A  and  Galetti D 2008 {\it Phys. Scr.} {\bf 78}
 045007.

 \bibitem{pati}   Loock P V and  Braunstein S L  2003 {\it Quantum Information with Continuous Variables}, eds.
Braunstein S L and Pati A K (Kluwer Academic, Dordrecht), P. 138.


\bibitem{[21]} Braunstein S L and Kimble H J 1998 {\it Phys. Rev. Lett.} {\bf 80} 869; Milburn
G J and Braunstein S L 1999 {\it Phys. Rev. A} {\bf 60} 937; Ralph
T C 2000 {\it Phys. Rev. A} {\bf 61} 010303(R).



\bibitem{suh}  Lee S-Y,  Qamar S,  Lee H-W and
 Zubairy M S 2008 {\it J. Phys. B: At. Mol. Opt. Phys.} {\bf 41}
 145504.

\bibitem{cont}
 Braunstein S L and  Loock P V  2005 {\it Rev.  Mod. Phy.}
{\bf  77}  513.

\bibitem{[24]}Dagenis M and Mandel L 1978 {\it Phys. Rev. A} {\bf 18} 2217.

\bibitem{flour} Carmichael H J and Walls D F 1976 {\it J. Phys.}
{\bf 9B} L43; Kimble H J and Mandel L 1976 {\it Phys. Rev. A} {\bf
13} 2123.
\bibitem{lou}  Loudon R 1980 {\it Rep. Prog. Phys.} {\bf 43} 913;  Reid M D and
 Walls D F 1986 {\it Phys. Rev. A} {\bf 34} 1260.

\bibitem{[3]}  Clauser J F 1974 {\it Phys. Rev. D} {\bf 9} 853.
\bibitem{[4]}  Kolchin P, Du S, Belthangady C, Yin G Y and Harris S
E 2006 {\it Phys. Rev. Lett.} {\bf 97 } 113602.
\bibitem{[5]}  Thompson J K, Simon J, Loh H and Vuletic V  2006 {\it Science} {\bf 313} 74.

\bibitem{[11]}  Li Y-q,  Edwards P J,  Huang  X and  Wang Y 2000 {\it J. Opt. B} {\bf 2} 292;
Marino A M, Boyer V and Lett P D quant-ph/0802.3183.
\bibitem{[27]} Agarwal G S 1988 {\it J. Opt. Soc. Am. B} {\bf 5}
1940.

\bibitem{carmich} Carmichael H J, Castro-Beltran H M, Foster G T and Orozco L A  2000 {\it Phys. Rev.
Lett.}  {\bf 85} 1855.

\bibitem{[34]} Banaszek K and Wódkiewicz k 1996 {\it Phys. Rev. Lett.} {\bf 76}
4344; Wallentowitz S and Vogel W 1996 {\it Phys. Rev. A} {\bf 53}
4528.
\bibitem{[35]}
Lutterbach L G and Davidovich L 1997 {\it Phys. Rev. Lett.} {\bf
78} 2547.
\bibitem{[36]}
Nogues G, Rauschenbeutel A, Osnaghi S, Bertet P, Brune M, Raimond
J M, Haroche S, Lutterbach L G and Davidovich L 2000 {\it Phys.
Rev. A} {\bf 62} 054101.

\bibitem{[37]} Beck M, Smithey D T and Raymer M G 1993 {\it Phys. Rev.
A} {\bf 48} 890; Smithey D T, Beck M, Cooper J and Raymer M G 1993
{\it Phys. Rev. A} {\bf 48} 3159; Beck M, Smithey D T, Cooper J
and Raymer M G 1993 {\it Opt. Lett.} {\bf 18} 1259; Smithey D T,
Beck M, Cooper J, Raymer M G and Faridani M B A 1993 {\it Phys.
Scr. T} {\bf 48} 35.


\bibitem{wolfgang} Schleich W P 2001 "Quantum optics in phase
space" (Wiley-VCH Verlag, Berlin)

\bibitem{faisalcoupler}  El-Orany F A A, Abdalla M S and  Pe\v{r}ina
J 2005 {\it Eur. Phys. J. D} {\bf 33} 453.

\bibitem{[48]} Leibfried D, Meekhof D M, King B E, Monroe C, Itano W M and
Wineland D J 1996 {\it Phys. Rev. Lett.} {\bf 77} 4281; Meekhof D
M, Monroe C, King B E, Itano W M and Wineland D J 1996 {\it Phys.
Rev. Lett.} {\bf 76} 1796; Monroe C, Meekhof D M, King B E and
Wineland D J 1996 {\it Science} {\bf 272} 1131.

\bibitem{haroche} Deleglise S, Dotsenko I, Sayrin C, Bernu
            J, Brune M, Raimond J-M and Haroche S 2008 {\it
            nature} {\bf 455} 510.
\end{thebibliography}
\end{document}